%% file: IGM8018.tex
\documentclass[twocolumn, tighten]{aastex7}

\usepackage{amsmath}

\shorttitle{Density Modulation of Ly$\alpha$}
\shortauthors{Zhu et al.}

\begin{document}

\title{Low Ly$\alpha$ Visibility in Galaxy Overdensities: Reionization Topology and Neutral-Fraction Ceilings from DIVER over $4.8<z<11$}

\author[0000-0003-3307-7525]{Yongda Zhu}
\thanks{JASPER Scholar}
\affiliation{Steward Observatory, University of Arizona, 933 North Cherry Avenue, Tucson, AZ 85721, USA}
\email[show]{yongdaz@arizona.edu}

\author[0000-0003-3310-0131]{Xiaohui Fan}
\affiliation{Steward Observatory, University of Arizona, 933 North Cherry Avenue, Tucson, AZ 85721, USA}
\email[]{xfan@arizona.edu}

\author[0000-0001-5211-1958]{Laura C. Keating}
\affiliation{Institute for Astronomy, University of Edinburgh, Blackford Hill, Edinburgh, EH9 3HJ, UK}
\email[]{Laura.Keating@ed.ac.uk}

\author[0000-0003-2344-263X]{George D. Becker}
\affiliation{Department of Physics \& Astronomy, University of California, Riverside, CA 92521, USA}
\email[]{georgeb@ucr.edu}

\author[0000-0003-1344-9475]{Eiichi Egami}
\affiliation{Steward Observatory, University of Arizona, 933 North Cherry Avenue, Tucson, AZ 85721, USA}
\email{egami@arizona.edu}

\author[0000-0001-6052-4234]{Xiaojing Lin}
\affiliation{Department of Astronomy, Tsinghua University, Beijing 100084, China}
\email{linxj21@mails.tsinghua.edu.cn}

\author[0000-0002-4622-6617]{Fengwu Sun}
\affiliation{Department of Astronomy, Westlake University, Hangzhou 310030, Zhejiang Province, China}
\affiliation{Center for Astrophysics $|$ Harvard \& Smithsonian, 60 Garden St., Cambridge, MA 02138, USA}
\email{sunfengwu@westlake.edu.cn}

\author[0000-0001-9420-7384]{Christopher Cain}
\affiliation{School of Earth and Space Exploration, Arizona State University, Tempe, AZ 85287-6004, USA}
\email{clcain3@asu.edu}

\author[0000-0002-7893-6170]{Marcia J. Rieke}
\affiliation{Steward Observatory, University of Arizona, 933 North Cherry Avenue, Tucson, AZ 85721, USA}
\email{mrieke@arizona.edu}

\author[0000-0002-8651-9879]{Andrew J.\ Bunker}
\affiliation{Department of Physics, University of Oxford, Denys Wilkinson Building, Keble Road, Oxford OX1 3RH, UK}
\email{andy.bunker@physics.ox.ac.uk}

\author[0009-0003-4133-0292]{Sijia Cai}
\affiliation{Department of Astronomy, Tsinghua University, Beijing 100084, China}
\email{caisj23@mails.tsinghua.edu.cn}

\author[0000-0003-2388-8172]{Francesco D'Eugenio}
\affiliation{Kavli Institute for Cosmology, University of Cambridge, Madingley Road, Cambridge, CB3 0HA, UK}
\affiliation{Cavendish Laboratory, University of Cambridge, 19 JJ Thomson Avenue, Cambridge, CB3 0HE, UK}
\email{fd391@cam.ac.uk}

\author[0000-0003-4337-6211]{Jakob M. Helton}
\affiliation{Department of Astronomy \& Astrophysics, The Pennsylvania State University, University Park, PA 16802, USA}
\email{jakobhelton@psu.edu}

\author[0000-0002-5768-738X]{Xiangyu Jin}
\affiliation{Department of Astronomy, University of Michigan, 1085 S. University Ave., Ann Arbor, MI 48109, USA}
\email{jxiangyu@umich.edu}

\author[0000-0001-6251-649X]{Mingyu Li}
\affiliation{Department of Astronomy, Tsinghua University, Beijing 100084, China}
\affiliation{Kavli Institute for Cosmology, University of Cambridge, Madingley Road, Cambridge, CB3 0HA, UK}
\affiliation{Cavendish Laboratory, University of Cambridge, 19 JJ Thomson Avenue, Cambridge, CB3 0HE, UK}
\email{lmy22@mails.tsinghua.edu.cn}

\author[0009-0003-5402-4809]{Zheng Ma}
\affiliation{Steward Observatory, University of Arizona, 933 North Cherry Avenue, Tucson, AZ 85721, USA}
\email{mazh@arizona.edu}

\author[0000-0002-4985-3819]{Roberto Maiolino}
\affiliation{Kavli Institute for Cosmology, University of Cambridge, Madingley Road, Cambridge, CB3 0HA, UK}
\affiliation{Cavendish Laboratory, University of Cambridge, 19 JJ Thomson Avenue, Cambridge, CB3 0HE, UK}
\affiliation{Department of Physics and Astronomy, University College London, Gower Street, London WC1E 6BT, UK}
\email{rm665@cam.ac.uk}

\author[0000-0002-5104-8245]{Pierluigi Rinaldi}
\affiliation{Space Telescope Science Institute, 3700 San Martin Drive, Baltimore, Maryland 21218, USA}
\email{prinaldi@stsci.edu}

\author[0000-0001-9262-9997]{Christopher N.\ A.\ Willmer}
\affiliation{Steward Observatory, University of Arizona, 933 North Cherry Avenue, Tucson, AZ 85721, USA}
\email{cnaw@arizona.edu}

\author[0000-0003-0111-8249]{Yunjing Wu}
\affiliation{Kavli Institute for the Physics and Mathematics of the Universe (WPI), The University of Tokyo Institutes for Advanced Study, The University of Tokyo, Kashiwa, Chiba 277-8583, Japan}
\affiliation{Center for Data-Driven Discovery, Kavli IPMU (WPI), UTIAS, The University of Tokyo, Kashiwa, Chiba 277-8583, Japan}
\email{yunjing.wu@ipmu.jp}

\author[0000-0002-8876-5248]{Zihao Wu}
\affiliation{Center for Astrophysics $|$ Harvard \& Smithsonian, 60 Garden St., Cambridge, MA 02138, USA}
\email{zihao.wu@cfa.harvard.edu}

\author[0000-0002-1574-2045]{Junyu Zhang}
\affiliation{Steward Observatory, University of Arizona, 933 North Cherry Avenue, Tucson, AZ 85721, USA}
\email{junyuzhang@arizona.edu}

\begin{abstract} 
Ly$\alpha$ emission is widely used to trace cosmic reionization, but its interpretation depends on how Ly$\alpha$ visibility varies with galaxy environment. We use deep JWST/NIRSpec observations from Deep Insights into UV Spectroscopy at the Epoch of Reionization (DIVER) in GOODS-N to measure Ly$\alpha$ visibility for 250 galaxies at $4.8<z<11$. The sample contains 84 Ly$\alpha$ detections, including 44 strong emitters with rest-frame Ly$\alpha$ equivalent width $W_{\rm Ly\alpha}>25$~\AA. We combine these measurements with H$\alpha$ and [O\,\textsc{iii}] emitters from JWST/NIRCam wide-field slitless spectroscopy to map the density field around each DIVER galaxy. Galaxies with high Ly$\alpha$ equivalent widths ($W_{\rm Ly\alpha}>25$~\AA) or high effective Ly$\alpha$ escape fractions ($f_{\rm esc,Ly\alpha}^{\rm eff}>0.05$) tend to lie farther from nearby H$\alpha$ and [O\,\textsc{iii}] emitters than galaxies with lower Ly$\alpha$ visibility. The clearest signal occurs near the prominent GOODS-N overdensity at $z\simeq5.2$, where fewer than 15\% of galaxies show strong Ly$\alpha$ emission. This trend is opposite to the simplest inside-out reionization expectation that overdensities produce larger ionized regions and enhance Ly$\alpha$ visibility. Possible explanations include circumgalactic and local intergalactic opacity, dense absorbers, and gas kinematics. We also derive an empirical upper envelope for $f_{\rm esc,Ly\alpha}^{\rm eff}$ and calibrate it with reionization simulations. Interpreting this envelope as a limiting IGM-attenuation signal gives neutral-fraction ceilings of $\langle x_{\rm HI}\rangle_{\rm max}=0.36$, 0.76, 0.74, 0.84, and 1.0 at $z\simeq5.2$, 5.8, 6.7, 7.7, and 9.8, respectively. The $z\sim8$ ceiling disfavors an almost completely neutral IGM at this epoch. These results support patchy reionization already underway by $z\sim8$ and show that galaxy Ly$\alpha$ visibility encodes both large-scale ionization topology and near-source gas structure. 
\end{abstract}


\section{Introduction} \label{sec:intro}

The timeline and topology of hydrogen reionization remain central to understanding the last major phase transition in the Universe. Ly$\alpha$ forest transmission toward $z\sim6$ quasi-stellar objects (QSOs) \citep[e.g.,][]{fan_constraining_2006,mcgreer_model-independent_2015} has long been interpreted as evidence that reionization was complete by $z\approx6$. Recent measurements, however, point to a process that ended late and was spatially inhomogeneous \citep[e.g.,][]{becker_evidence_2015,becker_evidence_2018,becker_damping_2024,bosman_new_2018,bosman_hydrogen_2022,mason_universe_2018,bolan_inferring_2022,yang_measurements_2020,zhu_chasing_2021,zhu_long_2022,whitler_insight_2023,zhu_damping_2024,spina_damping_2024,chen_impact_2026,umeda_probing_2026}.

The Ly$\alpha$ forest provides one of the clearest probes of the patchiness of reionization. At $z\sim5$--6, the effective Ly$\alpha$ optical depth, $\tau_{\rm eff}\equiv-\ln{\langle F\rangle}$, where $F$ is the continuum-normalized transmitted flux, shows large sightline-to-sightline scatter \citep[e.g.,][]{eilers_opacity_2018}. These fluctuations demonstrate that intergalactic medium (IGM) transmission varies on scales of tens of comoving megaparsecs \citep[cMpc; e.g.,][]{becker_evidence_2015,bosman_new_2018,yang_measurements_2020}. The scatter can arise from spatial variations in the ionizing ultraviolet background (UVB), residual neutral hydrogen, thermal relics of reionization, or a combination of these effects \citep{daloisio_large_2015,davies_large_2016,kulkarni_evolution_2019,nasir_observing_2020,keating_long_2020,garaldi_galaxy-igm_2024,garaldi_galaxy-igm_2025,zier_introducing_2026}. High-resolution radiative-hydrodynamic simulations further show that reionization-driven small-scale gas structure, self-shielding, and thermal evolution affect IGM opacity and Ly$\alpha$ transmission \citep{cain_introducing_2026}. These measurements constrain the timing and inhomogeneity of reionization, but they do not by themselves determine how Ly$\alpha$ transmission depends on galaxy environment.

Several studies have found that highly opaque Ly$\alpha$ troughs tend to lie in galaxy underdensities \citep[e.g.,][]{becker_evidence_2018,kashino_evidence_2020,christenson_constraints_2021,ishimoto_physical_2022}. Such a density-opacity relation arises naturally from a fluctuating UVB when the ionizing mean free path is short \citep[e.g.,][]{davies_determining_2018,becker_mean_2021,zhu_probing_2023,daloisio_large_2018}. Fluctuating-UVB models, including models in which neutral islands persist below $z\sim6$, are consistent with this picture \citep[e.g.,][]{kulkarni_evolution_2019,keating_long_2020,nasir_observing_2020}. Models in which the Ly$\alpha$ opacity scatter is driven mainly by residual temperature fluctuations have difficulty reproducing this relation \citep[e.g.,][]{daloisio_large_2015}, because recently reionized low-density regions should remain hot and relatively transmissive. UVB fluctuations, temperature fluctuations, and residual neutral structures may all contribute to the observed density-opacity relation.

The relation is not one-to-one. \citet{christenson_relationship_2023} found that two highly transmissive Ly$\alpha$ forest sightlines near the end of reionization ($z\sim5.7$), toward the QSOs SDSS J1306+0356 and PSOJ359$-$06, are located in galaxy underdensities traced by Ly$\alpha$ emitters (LAEs), showing that low IGM Ly$\alpha$ opacity does not always coincide with galaxy overdensity. Follow-up modeling by \citet{gangolli_correlation_2025} using the FlexRT simulations \citep{cain_flexrt_2024} showed that transmissive underdense sightlines can occur if voids were reionized relatively late and remain hot, while transmissive overdense regions can arise from locally enhanced ionizing backgrounds. \citet{zhu_galaxy_2026} used JWST [O\,\textsc{iii}] emitters to revisit these highly transmissive QSO sightlines and confirmed that the clearest IGM Ly$\alpha$ transmission at $z=5.7$ occurs in galaxy underdensities. This result shows that the trend identified by \citet{christenson_relationship_2023} is not driven solely by LAE selection effects. Galaxy--Ly$\alpha$-forest cross-correlations likewise show that absorption and transmission around galaxies can vary over several to tens of comoving megaparsecs \citep{kakiichi_jwst_2025,kashino_eiger_2025}.

Ly$\alpha$ emission from galaxies provides a complementary probe by linking IGM transmission to the sources themselves. Because Ly$\alpha$ photons are resonantly scattered, the observed line strength depends on radiative transfer through the interstellar medium (ISM), circumgalactic medium (CGM), and IGM \citep[e.g.,][]{mason_universe_2018,bolan_inferring_2022,hashemi_ly_2025}. The redshift evolution of the fraction of galaxies with strong Ly$\alpha$ emission has therefore been used to constrain the IGM neutral fraction across reionization \citep[e.g.,][]{stark_keck_2010,schenker_line-emitting_2014,tang_jwstnirspec_2023,tang_jwstnirspec_2024,tang_ly_2024,kageura_census_2025,chen_impact_2026}. In a simple inside-out reionization picture, galaxy overdensities should host larger ionized regions and stronger Ly$\alpha$ visibility. This expectation need not hold on the scales probed by galaxy Ly$\alpha$: optically thick absorbers and residual neutral structures can suppress Ly$\alpha$ transmission even within substantially ionized regions \citep{bolton_rapid_2013,mesinger_can_2015}, while neutral-gas covering in the CGM and gas kinematics can further alter the emergent line.

JWST/NIRSpec surveys now provide population measurements of Ly$\alpha$ visibility across broad redshift ranges \citep{jones_jades_2024,jones_jades_2025}. JADES studies have also identified an extreme-equivalent-width LAE at $z=7.3$ and a sample of faint LAEs at $z\simeq5.8$--8 with measured Ly$\alpha$ escape and ionizing properties. Environmental analyses associate some, but not all, of these LAEs with overdensities or inferred ionized bubbles, while detections at $z>8$ require favorable ionized paths \citep{saxena_jades_2023,saxena_jades_2024,witstok_inside_2024,witstok_jades_2025}. The presence of strong LAEs in overdensities does not by itself imply that overdense environments have higher Ly$\alpha$ visibility. These studies identify galaxies with detected Ly$\alpha$ emission but generally do not measure the visibility distribution of all galaxies in the same structures. Other JWST studies have found substantial field-to-field variation in Ly$\alpha$ visibility and candidate ionized regions associated with galaxy overdensities \citep{napolitano_peering_2024,napolitano_ly_2026}.

The environmental dependence of galaxy Ly$\alpha$ visibility remains poorly constrained, limiting its use as a direct probe of the reionization timeline and topology. In GOODS-N, the DIVER program (PID 8018; PI: X.~Lin) provides JWST/NIRSpec Ly$\alpha$ coverage for hundreds of galaxies across $4.8<z<11$, spanning a wide redshift interval and diverse galaxy environments. The First Reionization Epoch Spectroscopically Complete Observations survey (FRESCO; \citealp{oesch_jwst_2023,meyer_jwst_2024,covelo-paz_h_2025}) and the Complete NIRCam Grism Redshift Survey (CONGRESS; \citealp{lin_luminosity_2025}; F.~Sun et al.\ in prep.) provide a complementary JWST/NIRCam wide-field slitless spectroscopy (WFSS) redshift catalog that traces the density field with H$\alpha$ and [O\,\textsc{iii}] emitters selected from rest-frame optical lines. We combine the DIVER Ly$\alpha$ measurements with the WFSS tracer field to test whether the most Ly$\alpha$-visible galaxies occupy the most overdense line-emitter environments. The broad redshift coverage also allows us to use the upper envelope of Ly$\alpha$ visibility to place neutral-fraction ceilings across reionization.

\begin{figure*}[!ht]
    \centering
    \includegraphics[width=\textwidth]{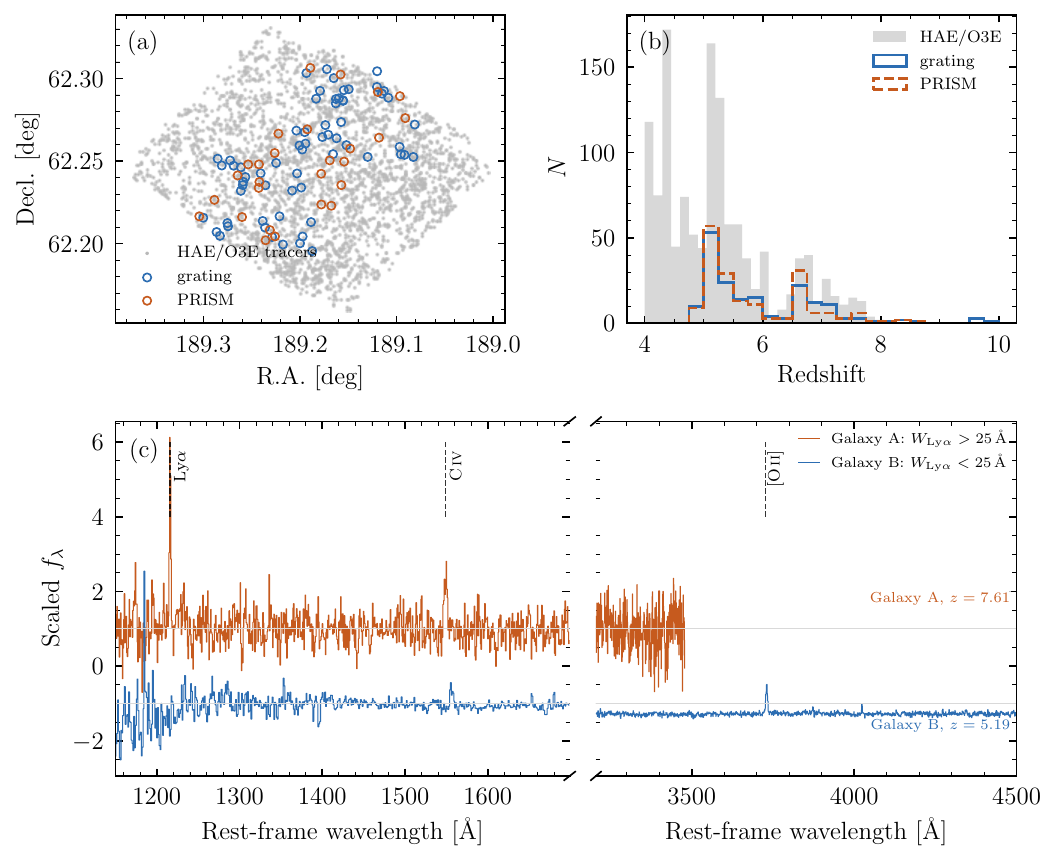}
    \caption{
    Overview of the DIVER data set and environmental measurements used in this work. \textbf{(a)}: GOODS-N footprint of the WFSS line-emitter tracer catalog and the JWST/NIRSpec grating and PRISM samples. \textbf{(b)}: Redshift distributions of the WFSS tracer catalog and DIVER science sample. \textbf{(c)}: Example grating spectra. Galaxy A at $z=7.61$ shows strong Ly$\alpha$ emission, while Galaxy B at lower redshift shows no detected Ly$\alpha$ emission. This paper tests whether such differences in Ly$\alpha$ visibility are linked to the surrounding galaxy density field.
    }
    \label{fig:overview}
\end{figure*}

This paper is organized as follows. Section~\ref{sec:data} describes the Ly$\alpha$ spectroscopy and density-field construction. Section~\ref{sec:results} presents the redshift and environmental dependence of Ly$\alpha$ visibility. Section~\ref{sec:discussion} discusses the implications for reionization topology and compares the measurements with recent simulations. Section~\ref{sec:discussion_ceiling} presents the neutral-fraction ceiling method and its constraints on the reionization timeline. Section~\ref{sec:summary} summarizes the conclusions. We adopt the Planck flat $\Lambda$CDM cosmology, $H_0=67.4$ km s$^{-1}$ Mpc$^{-1}$, $\Omega_{\rm m}=0.315$, and $\Omega_\Lambda=0.685$, with $h\equiv H_0/(100~{\rm km~s^{-1}~Mpc^{-1}})=0.674$ \citep{planck_collaboration_planck_2020}. Comoving distances are used unless otherwise stated. Figure~\ref{fig:overview} summarizes the DIVER measurements and tracer sample.

\section{Data and Measurements} \label{sec:data}

The DIVER observations contain 283 unique sources. The median accumulated exposure times of the extracted spectra are 20.4~hr in the F070LP/G140M grating configuration ($R\simeq1000$) and 2.3~hr in PRISM ($R\simeq100$). The extracted spectra cover approximately $0.70$--$2.91~\mu$m for G140M and $0.60$--$5.30~\mu$m for PRISM. The Ly$\alpha$ search over $4.8<z<11$ uses the wavelength range $0.70$--$1.46~\mu$m.

The DIVER targets are mainly grism-selected emission-line galaxies from FRESCO and CONGRESS, together with active galactic nuclei (AGN), little red dots, and high-redshift photometric candidates (X.~Lin et al., in prep.). The microshutter assembly (MSA) target lists were ranked by catalog membership and priority, without prior knowledge of Ly$\alpha$ visibility or the surrounding galaxy density. DIVER used separate MSA designs for the grating and PRISM observations: two G140M configurations in observation 1 and five PRISM configurations across observations 2 and 3. The target lists and shutter assignments differed between the two modes, and detector gaps, wavelength coverage, masked pixels, and spectral-quality cuts further affected the available Ly$\alpha$ coverage. The resulting grating and PRISM samples overlap but are not identical. We do not add JADES/NIRSpec spectra \citep{bunker_jades_2024,deugenio_jades_2025,curtis-lake_jades_2025,scholtz_jades_2025} because the available G140M coverage is not uniform or comparable in depth across the JADES sample.

\subsection{NIRSpec data reduction}

The NIRSpec reduction follows the method used in \citet{zhu_smiles_2026}, adapted for the DIVER observations. We reduce the raw exposures with the JWST Calibration Pipeline v1.20.2 \citep{bushouse_jwst_2025} and the Calibration Reference Data System \footnote{\url{https://jwst-crds.stsci.edu}} reference file \texttt{jwst\_1464.pmap}, using custom scripts throughout the reduction.

We begin with the \texttt{uncal.fits} exposures and run \texttt{calwebb\_detector1} with revised jump and snowball treatments to produce the \texttt{rate.fits} files. We then remove hot pixels and $1/f$ noise before running \texttt{calwebb\_spec2} for each exposure and nod position.

We apply two additional steps during the slitlet-level reduction. First, we extend the wavelength calibration of the F070LP/G140M spectra to $\simeq3~\mu$m following \citet{scholtz_jades_2025}. Second, in regions with densely packed slit traces, we apply a master-sky subtraction to reduce contamination from overlapping spectra, diffuse zodiacal emission, and second- and third-order light at $\gtrsim1.3~\mu$m. For each slitlet, the master sky is constructed from the median of the available exposures and subtracted from the science array. We retain a version without master-sky subtraction to check that the procedure does not introduce spectral features. The main measurements use the standard three-nod background-subtracted products for most sources. We then run \texttt{calwebb\_spec3} to combine the exposures and extract the final spectra.

Appendix~\ref{app:data_validation} presents the source catalog. It also compares the grating, PRISM, SED, and WFSS measurements (see below).

\subsection{\texorpdfstring{Ly$\alpha$ measurements}{Lyalpha measurements}} \label{sec:data_lya}

We measure Ly$\alpha$ primarily from the F070LP/G140M spectra. PRISM measurements are used when grating coverage is unavailable and for consistency checks between the two modes. The expected Ly$\alpha$ wavelength is calculated from the adopted systemic redshift.

For PRISM sources, we adopt a rest-frame optical redshift measured from the PRISM spectrum when H$\alpha$ or [O\,\textsc{iii}] $\lambda5008$ is detected at ${\rm S/N}>3$. The redshift is checked against [O\,\textsc{iii}] $\lambda4960$ and H$\beta$ when available. Otherwise, we retain the input spectroscopic catalog redshift. For grating-only sources, the catalog redshift is based mainly on CONGRESS or FRESCO H$\alpha$ and [O\,\textsc{iii}] measurements, with existing NIRSpec redshifts used where available. We revise the adopted redshift after manual inspection when needed. We exclude spectra in which the expected Ly$\alpha$ region is strongly affected by detector gaps, incomplete wavelength coverage, or poor spectral quality.

The grating continuum is generally too weak to determine the rest-UV continuum used for the equivalent-width measurements. The local continuum fit is used only to remove the spectral baseline. We define a window around the expected Ly$\alpha$ wavelength and mask the line region over velocities from $-700$ to $+1400$~km~s$^{-1}$ relative to the expected line center. We fit the remaining sidebands with a low-order spline, using a linear or constant model when the sideband coverage is insufficient.

We measure Ly$\alpha$ with both a single-Gaussian fit and direct summation of the continuum-subtracted spectrum. The Gaussian fit gives the line center, width, and integrated flux. Some well-resolved Ly$\alpha$ lines show asymmetric profiles, but we use the same single-Gaussian model for all sources to maintain a uniform measurement procedure. Direct summation measures the flux over the fitted line interval without assuming a line profile. We adopt the summed flux when available and use the Gaussian flux otherwise. Direct summation is unavailable when fewer than two unmasked pixels with positive uncertainties lie within the integration interval, as can occur near a masked region or the edge of the spectral coverage. In these cases, the wider fitting window can still provide a Gaussian measurement.

The Ly$\alpha$ luminosity is calculated from the adopted line flux and redshift. For each source, we inspect local continuum fit, continuum-subtracted spectrum, and Gaussian model. For non-detections with usable wavelength coverage, we retain the 3$\sigma$ upper limits on the line flux and rest-frame equivalent width. These galaxies are included in the non-strong Ly$\alpha$ population.

\subsection{Line-emitter density proxy} \label{sec:data_density}

We trace galaxy environment with a combined JWST/NIRCam wide-field slitless spectroscopy (WFSS) catalog of H$\alpha$ emitters (HAEs) and [O\,\textsc{iii}] emitters (O3Es) in GOODS-N (F.~Sun et al., in prep.). The catalog combines data from CONGRESS (JWST-GO-3577; PI: E. Egami; DOI: 10.17909/6rfk-6s81; \citealp{lin_luminosity_2026}) and FRESCO (JWST-GO-1895; PI: P. Oesch; DOI: 10.17909/gdyc-7g80; \citealp{oesch_jwst_2023,meyer_jwst_2024}). The underlying photometry comes from the JADES DR5 catalog \citep{robertson_jwst_2026}. The line-fitting procedure and detection criteria follow the SAPPHIRES Early Data Release \citep{sun_slitless_2025}. Because the tracers are selected through rest-frame optical emission lines, their selection does not depend directly on Ly$\alpha$ transmission.

FRESCO provides F444W grism coverage over approximately $3.8$--$5.0~\mu$m, selecting O3Es at $z\simeq6.6$--9.0 and HAEs at $z\simeq4.8$--6.6. CONGRESS provides F356W grism coverage over approximately $3.1$--$4.0~\mu$m, selecting O3Es at $z\simeq5.2$--7.0 and HAEs at $z\simeq3.7$--5.1. Over the DIVER redshift range, HAEs trace the density field at the low-redshift end, while O3Es provide most of the coverage over $z\simeq5$--9. The tracer coverage is strongest at $z\simeq5$--8.

The catalog includes line-spread-function-aided fits for candidate lines with ${\rm S/N}>3$. Inclusion in the catalog requires at least one line with ${\rm S/N}>4$ and a quadrature-combined significance greater than 5 for lines with ${\rm S/N}>3$. We use sources with $\texttt{flag\_DR}=1$ and require $\texttt{zbest\_conf}\geq3$. The latter requires a line detection with ${\rm S/N}>3$ and agreement between the photometric and cross-correlation redshifts,
\begin{equation}
|z_{\rm phot}-z_{\rm xcorr}|<0.2,
\end{equation}
where $z_{\rm xcorr}$ is obtained by cross-correlating the detected rest-frame optical lines with line templates. Because the sensitivity varies with filter, wavelength, and line species, the tracer sample does not have a single flux limit. We also caution that it is an emission-line-selected sample and is not complete in stellar or halo mass.

Our main environment statistic is the normalized neighbor distance,
\begin{equation}
\frac{d_{\rm mean3}}{d_{\rm mean3,rand}(z)}.
\end{equation}
For each DIVER galaxy, $d_{\rm mean3}$ is the mean projected comoving distance to the nearest three WFSS tracers within a line-of-sight interval of $\pm10$~cMpc. We exclude self-matches within 0\farcs3. The reference value $d_{\rm mean3,rand}(z)$ is calculated from 500 random positions within the convex hull of the WFSS footprint in each $\Delta z=0.1$ bin, using the same line-of-sight interval. The random reference accounts for the survey footprint and the redshift-dependent abundance of the selected tracers. One caveat is that it does not correct for spatial variations in WFSS sensitivity or contamination. Appendix~\ref{app:visibility_robustness} presents tests using alternative neighbor numbers and randomized tracer positions.

The nearest-three statistic balances locality and shot noise in the WFSS tracer field. The visibility--environment ordering remains unchanged when the mean distance is calculated from the nearest five or seven tracers. Tests using different line-of-sight intervals also give the same qualitative relation. For a locally uniform projected tracer distribution, the nearest-neighbor distance scales approximately as
\begin{equation}
d\propto\Sigma^{-1/2}.
\end{equation}
The corresponding surface-density ratio is
\begin{equation}
\frac{\Sigma}{\Sigma_{\rm rand}}
\simeq
\left[
\frac{d_{\rm mean3}}{d_{\rm mean3,rand}}
\right]^{-2}.
\end{equation}
The normalized neighbor distance is therefore a good monotonic inverse proxy for projected line-emitter density \citep[e.g.,][]{christenson_relationship_2023,gangolli_correlation_2025,zhu_galaxy_2026}.

\subsection{\texorpdfstring{Ly$\alpha$ visibility quantities}{Lyalpha visibility quantities}} \label{sec:data_sed}

We combine the Ly$\alpha$ measurements with the rest-UV continuum and H$\alpha$ luminosity to calculate the Ly$\alpha$ equivalent width and effective Ly$\alpha$ visibility. The grating spectra generally do not provide H$\alpha$ or a uniform measurement of the rest-UV continuum. For the grating sample, we therefore use photometry-based spectral energy distribution (SED) fits based on deep JADES imaging \citep{eisenstein_overview_2023,rieke_jades_2023,johnson_jwst_2026,robertson_jwst_2026,wu_jwst_2026}.

The photometry includes JWST/NIRCam imaging in F090W, F115W, F150W, F162M, F182M, F200W, F210M, F277W, F335M, F356W, F410M, and F444W, together with Hubble Space Telescope imaging in F435W, F606W, F775W, F814W, and F850LP \citep{whitaker_hubble_2019}. We fit the photometry with \texttt{Prospector} \citep{johnson_stellar_2021}, following \citet{zhu_smiles_2026}. The models include non-parametric star-formation histories \citep{leja_how_2019}, dust attenuation, and nebular emission \citep{leja_deriving_2017}.

The SED fits are mainly used to estimate the continuum near Ly$\alpha$ and the H$\alpha$ flux required for the visibility measurements. They are not used to predict the escaping Ly$\alpha$ emission. The quantities used here are the rest-frame continuum near Ly$\alpha$, the observed UV absolute magnitude $M_{1500,\rm obs}$, and the observed H$\alpha$ luminosity $L_{{\rm H}\alpha,\rm obs}$. Photometric H$\alpha$ inference has also been tested for low-mass galaxies \citep{simmonds_low-mass_2024}.

For PRISM sources, we use the UV continuum and H$\alpha$ luminosity measured from the PRISM spectrum. For the 147 sources shown in Appendix~\ref{app:data_validation}, the median $\log_{10}(F_{\rm SED}/F_{\rm PRISM})$ is $+0.064$ dex, with a scatter of 0.223 dex calculated as 1.4826 times the median absolute deviation. The small median offset shows that the two measurements are broadly consistent, while the scatter reflects substantial source-to-source differences. For 110 sources with WFSS measurements satisfying the spectral-quality criteria, the corresponding comparison gives a median $\log_{10}(F_{\rm SED}/F_{\rm WFSS})=+0.072$ dex and a scatter of 0.170 dex.

We pair PRISM Ly$\alpha$ with the PRISM H$\alpha$ measurement and grating Ly$\alpha$ with the SED-derived H$\alpha$ measurement. The WFSS H$\alpha$ measurements are used only for validation. We define
\begin{equation}
f_{\rm esc,Ly\alpha}^{\rm eff}
\equiv
\frac{L_{\rm Ly\alpha}}
{8.7\,L_{{\rm H}\alpha,\rm obs}},
\end{equation}
where 8.7 is the Case B Ly$\alpha$/H$\alpha$ luminosity ratio \citep{osterbrock_astrophysics_2006}. The observed H$\alpha$ luminosity is not corrected for dust attenuation.

Ignoring departures from the Case B ratio, this quantity can be written schematically as
\begin{equation}
f_{\rm esc,Ly\alpha}^{\rm eff}
\sim
f_{\rm Ly\alpha}^{\rm ISM}\,
T_{\rm CGM}\,
T_{\rm IGM}\,
\frac{a_{\rm Ly\alpha}}{a_{{\rm H}\alpha}}\,
10^{0.4A_{{\rm H}\alpha}},
\end{equation}
where $f_{\rm Ly\alpha}^{\rm ISM}$ is the fraction of Ly$\alpha$ transmitted through the ISM, $T_{\rm CGM}$ and $T_{\rm IGM}$ are the CGM and IGM transmission factors, $a_{\rm Ly\alpha}/a_{{\rm H}\alpha}$ is the relative aperture recovery, and $A_{{\rm H}\alpha}$ is the attenuation of the observed H$\alpha$ emission.

Attenuation in the ISM, CGM, and IGM and preferential Ly$\alpha$ slit loss lower $f_{\rm esc,Ly\alpha}^{\rm eff}$. Dust attenuation or aperture loss in H$\alpha$ lowers the denominator and raises the ratio. An overestimated H$\alpha$ luminosity lowers the ratio, and an underestimated one raises it. We therefore use $f_{\rm esc,Ly\alpha}^{\rm eff}$ as an effective visibility statistic that includes radiative-transfer, aperture, and H$\alpha$-normalization effects.

The grating and PRISM spectra have different Ly$\alpha$ sensitivities. The median 3$\sigma$ line-flux limits are $1.17\times10^{-18}$ and $3.65\times10^{-18}$~erg~s$^{-1}$~cm$^{-2}$ for the grating and PRISM measurements, respectively. The corresponding median rest-frame equivalent-width limits are 15.6 and 74.4~\AA. We therefore rely mainly on the deeper grating sample for the environmental comparisons and treat the grating and PRISM measurements separately where needed. Each Ly$\alpha$ measurement is paired with the continuum and H$\alpha$ measurement from the corresponding method.

One caveat is that the grating Ly$\alpha$ fluxes and the photometry-based continuum and H$\alpha$ measurements do not use identical apertures. Wavelength-dependent slit losses, source morphology, and spatial offsets between Ly$\alpha$ and the rest-UV continuum can bias $W_{\rm Ly\alpha}$ and $f_{\rm esc,Ly\alpha}^{\rm eff}$ for individual sources. \citet{napolitano_ly_2026} found an average Ly$\alpha$ slit loss of $35\pm10\%$ by comparing JWST/NIRSpec and VLT/FORS2 measurements at $z\simeq6$. We do not apply an explicit slit-loss correction. However, all grating sources are reduced, extracted, and measured in the same way, which preserves a uniform scale for the relative comparisons. The grating--PRISM and PRISM--SED comparisons in Appendix~\ref{app:data_validation} also show no strong global offset.

\subsection{Final analysis samples} \label{sec:data_sample}

We exclude the 7 broad-line/type-1 AGN identified by \citet{zhang_abundant_2026}, because AGN emission can alter both the intrinsic Ly$\alpha$/H$\alpha$ ratio and the SED-based H$\alpha$ normalization. After applying the redshift, wavelength-coverage, detector-gap, edge-contamination, and spectral-quality criteria, the spectroscopic parent sample contains 262 galaxies. Of these, 166 have usable grating coverage, 163 have usable PRISM coverage, and 67 have both. For galaxies observed with both modes, we adopt the grating measurement.

One galaxy has ambiguous JADES photometry because of an uncertain source match and blended segmentation, preventing a reliable UV-continuum or H$\alpha$ measurement. It remains in the 262-galaxy spectroscopic parent sample but is excluded from the 261-source machine-readable table and from analyses requiring an equivalent width or $f_{\rm esc,Ly\alpha}^{\rm eff}$. The EW-classification sample contains 250 galaxies, including 44 strong and 206 non-strong Ly$\alpha$ emitters. Of these, 179 have environment measurements, and 177 also have the normalized neighbor distance used in the $f_{\rm esc,Ly\alpha}^{\rm eff}$--environment analysis. The density coverage becomes less complete at the highest redshifts, where the WFSS tracer sample is sparse.

We define strong Ly$\alpha$ emitters as galaxies with rest-frame equivalent width $W_{\rm Ly\alpha}>25$~\AA\ measured at EW ${\rm S/N}>3$. Table~\ref{tab:sample_summary} summarizes the available measurements and analysis samples.

\begin{deluxetable*}{ccccccccccc}
\tabletypesize{\scriptsize}
\setlength{\tabcolsep}{2.5pt}
\tablewidth{0pt}
\tablecaption{Sample-size summary\label{tab:sample_summary}}
\tablehead{
\colhead{Grating} &
\colhead{PRISM} &
\colhead{Photometry} &
\colhead{Environment} &
\colhead{\shortstack{Ly$\alpha$\\detection}} &
\colhead{\shortstack{Strong\\Ly$\alpha$}} &
\colhead{\shortstack{Non-strong\\Ly$\alpha$}} &
\colhead{\shortstack{Ly$\alpha$ visibility\\+ environment}} &
\colhead{\shortstack{EW class\\+ environment}} &
\colhead{\shortstack{Upper-envelope\\sample}} &
\colhead{Count}
}
\startdata
x &  & x & x &  &  & x & x & x &  & 53 \\
 & x & x &  &  &  & x &  &  &  & 46 \\
x & x & x & x &  &  & x & x & x &  & 23 \\
x & x & x & x & x & x &  & x & x & x & 20 \\
 & x & x & x &  &  & x & x & x &  & 19 \\
x &  & x &  &  &  & x &  &  &  & 15 \\
 & x & x & x & x &  & x & x & x & x & 15 \\
x & x & x & x & x &  & x & x & x & x & 14 \\
x &  & x & x & x & x &  & x & x & x & 13 \\
x &  & x & x & x &  & x & x & x & x & 10 \\
 & x & x & x & x & x &  & x & x & x & 10 \\
x & x & x &  &  &  & x &  &  &  & 8 \\
 & x & x &  &  &  &  &  &  &  & 5 \\
x &  & x &  &  &  &  &  &  &  & 4 \\
x &  & x &  &  &  &  &  &  & x & 2 \\
x & x & x & x &  &  & x &  & x &  & 1 \\
x & x & x &  & x & x &  &  &  & x & 1 \\
x &  & x &  & x &  & x &  &  & x & 1 \\
x &  &  &  &  &  &  &  &  &  & 1 \\
 & x & x & x &  &  & x &  & x &  & 1 \\
\tableline
\textbf{166} & \textbf{163} & \textbf{261} & \textbf{179} & \textbf{84} & \textbf{44} & \textbf{206} & \textbf{177} & \textbf{179} & \textbf{86} & \shortstack{\textbf{Parent sample}\\\textbf{262}} \\
\enddata
\tablecomments{Each row represents one combination of measurement availability and analysis-sample membership within the 262-galaxy spectroscopic parent sample. An ``x'' indicates that the corresponding measurement is available or that the galaxy belongs to the indicated sample; a blank entry indicates otherwise. The Count column gives the number of galaxies with each combination. The final row gives the total number of galaxies in each column. Because the samples overlap, the column totals should not be added together.}
\end{deluxetable*}

We characterize Ly$\alpha$ visibility in two ways. First, we calculate the fraction of galaxies with $W_{\rm Ly\alpha}>25$~\AA. Non-detections are included in the denominator when the expected Ly$\alpha$ wavelength has usable spectral coverage. Second, we use the continuous quantity $f_{\rm esc,Ly\alpha}^{\rm eff}$, which measures the Ly$\alpha$ luminosity relative to the inferred H$\alpha$ luminosity. The grating measurements use the SED-based UV continuum and H$\alpha$ luminosity, while the PRISM measurements use the corresponding spectral measurements. The upper-envelope analysis uses the grating Ly$\alpha$ detection when available and the PRISM detection otherwise.

For a simple high- and low-visibility comparison, we divide the deeper grating sample at $f_{\rm esc,Ly\alpha}^{\rm eff}=0.05$. This threshold corresponds to a Ly$\alpha$ luminosity equal to 5\% of the Case B expectation based on the observed H$\alpha$ luminosity. We estimate the distribution with the Kaplan--Meier method, which combines measured values with upper limits instead of treating the limits as detections. The resulting median for the full grating sample is 0.047, close to the adopted threshold. Sources detected above 0.05 are classified as high visibility. Detections or constraining 3$\sigma$ upper limits at or below 0.05 are classified as low visibility, while upper limits above 0.05 remain unclassified. We do not use the shallower PRISM measurements for this binary comparison because most PRISM upper limits do not reach $f_{\rm esc,Ly\alpha}^{\rm eff}=0.05$.

As a complementary test, we compare the full $f_{\rm esc,Ly\alpha}^{\rm eff}$ distributions while retaining the upper limits. The Kaplan--Meier method estimates each distribution, and the log-rank test evaluates whether the two distributions differ after accounting for the upper limits. The grating and PRISM measurements are treated separately. We also use a permutation test to estimate how often the observed environmental separation could arise by chance. In this test, we randomly reassign the measured environments among galaxies and recalculate the separation, while keeping the grating and PRISM samples separate. The probability is given by the fraction of randomized samples with a separation at least as large as the observed one. For comparison with the Keating et al.\ simulations (in prep.; hereafter K26), we divide the grating sample within each observational redshift interval at its Kaplan--Meier median. The K26 galaxies are divided at the median $f_{\rm esc,IGM}$ of each simulation snapshot. These median divisions are used only for the simulation comparison.

\begin{figure*}[!ht]
    \centering
    \includegraphics[width=\textwidth]{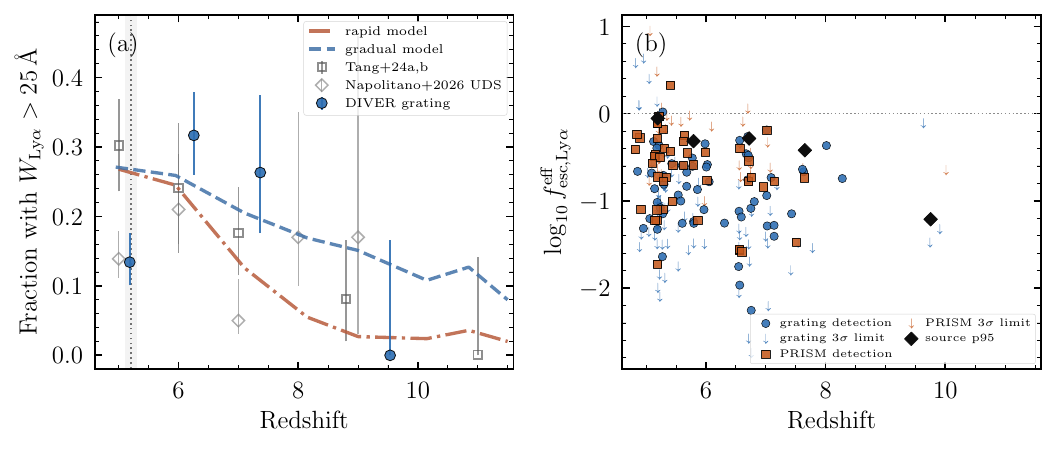}
    \caption{
    Redshift evolution of Ly$\alpha$ visibility. \textbf{(a)} Fraction of the 166 grating sources with $W_{\rm Ly\alpha}>25$~\AA\ at EW ${\rm S/N}>3$. Literature measurements using the same EW threshold are shown together with the K26 rapid and gradual reionization models (Keating et al., in prep.). The dotted vertical line in panel~(a) marks $z=5.2$, the center of the highlighted GOODS-N overdensity interval. \textbf{(b)} Effective Ly$\alpha$ visibility for the grating and PRISM samples. Filled symbols show 59 grating and 47 PRISM detections, and downward arrows show 81 grating and 34 PRISM 3$\sigma$ upper limits. Black diamonds mark the 95th-percentile upper envelope calculated from 86 galaxies with Ly$\alpha$ detections; 84 of these have environment measurements.
    }
    \label{fig:lya_fraction_redshift}
\end{figure*}

\begin{figure*}[!ht]
    \centering
    \includegraphics[width=\textwidth]{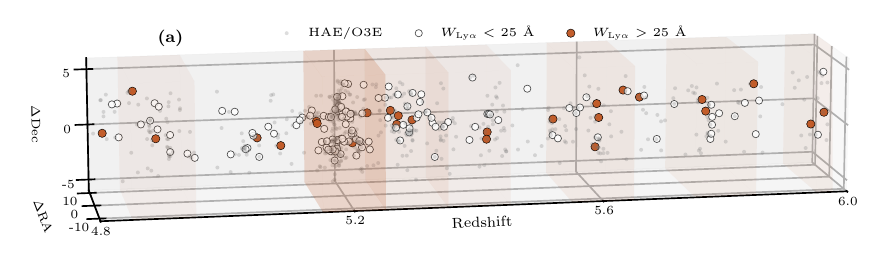}
    \includegraphics[width=\textwidth]{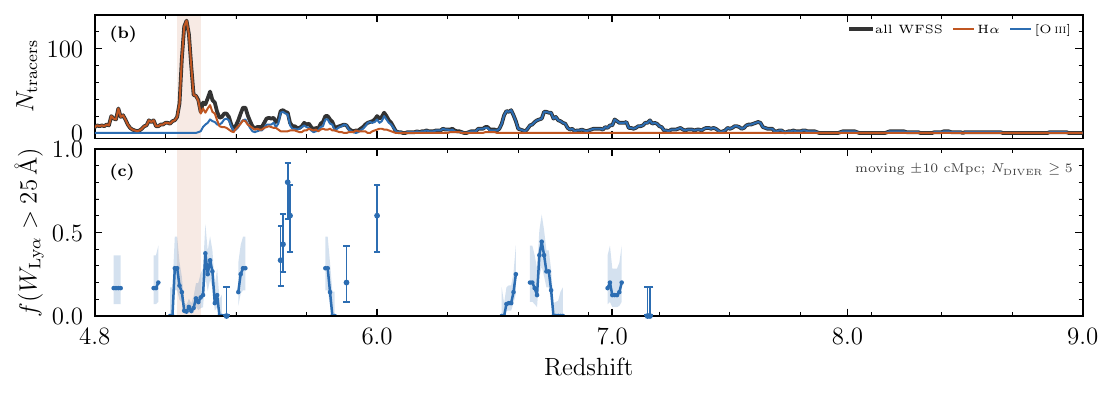}
    \caption{
    Ly$\alpha$ visibility across the GOODS-N density field. \textbf{(a)} Redshift-space distribution of the WFSS line emitters and DIVER galaxies over $4.8<z<6$. Gray points show HAEs and O3Es, open circles show galaxies with $W_{\rm Ly\alpha}<25$~\AA, and orange circles show galaxies with $W_{\rm Ly\alpha}>25$~\AA. \textbf{(b)} Number of all WFSS tracers, HAEs, and O3Es within a moving $\pm10$~cMpc window. \textbf{(c)} Strong-Ly$\alpha$ fraction in the same window. Shaded regions show the 68\% binomial intervals, and windows containing fewer than five DIVER galaxies are omitted. The vertical shading marks the $z\simeq5.2$ overdensity over $5.15<z<5.25$.
    }
    \label{fig:overdensity_suppression}
\end{figure*}

\begin{figure*}[!ht]
    \centering
    \includegraphics[width=\textwidth]{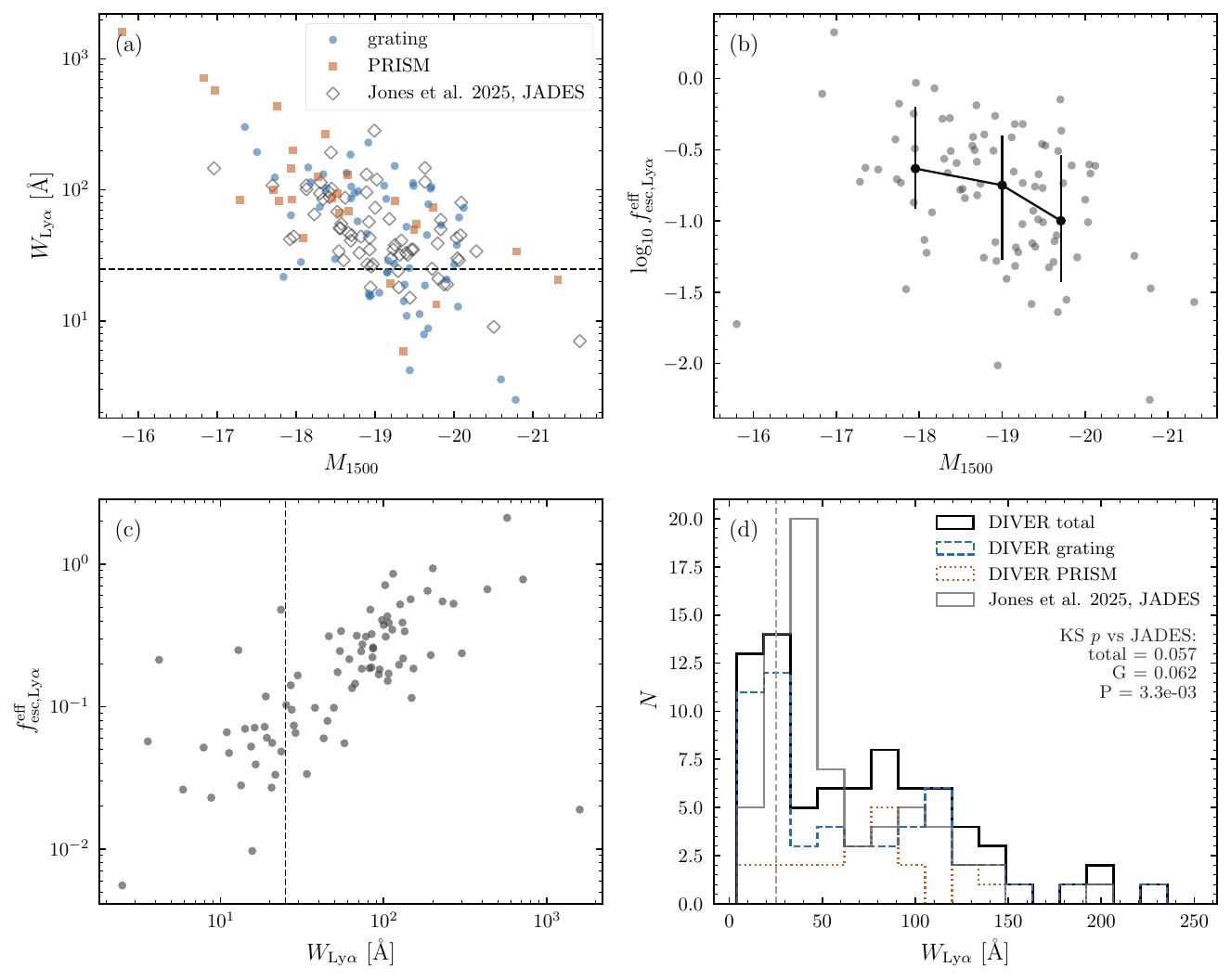}
    \caption{
    Ly$\alpha$ equivalent width and effective visibility. \textbf{(a)} Ly$\alpha$ EW as a function of $M_{1500,\rm obs}$, compared with the detected JADES LAEs from \citet{jones_jades_2025}. The horizontal dashed line marks $W_{\rm Ly\alpha}=25$~\AA. \textbf{(b)} Effective Ly$\alpha$ visibility as a function of $M_{1500,\rm obs}$. Black points show the measurements in UV-luminosity bins. \textbf{(c)} Effective Ly$\alpha$ visibility as a function of Ly$\alpha$ EW. The vertical dashed line marks $W_{\rm Ly\alpha}=25$~\AA. \textbf{(d)} Ly$\alpha$ EW distributions for the total, grating, and PRISM samples, compared with the JADES LAE distribution.
    }
    \label{fig:lya_properties_m1500}
\end{figure*}

\begin{figure*}[!ht]
    \centering
    \includegraphics[width=\textwidth,height=0.70\textheight,keepaspectratio]{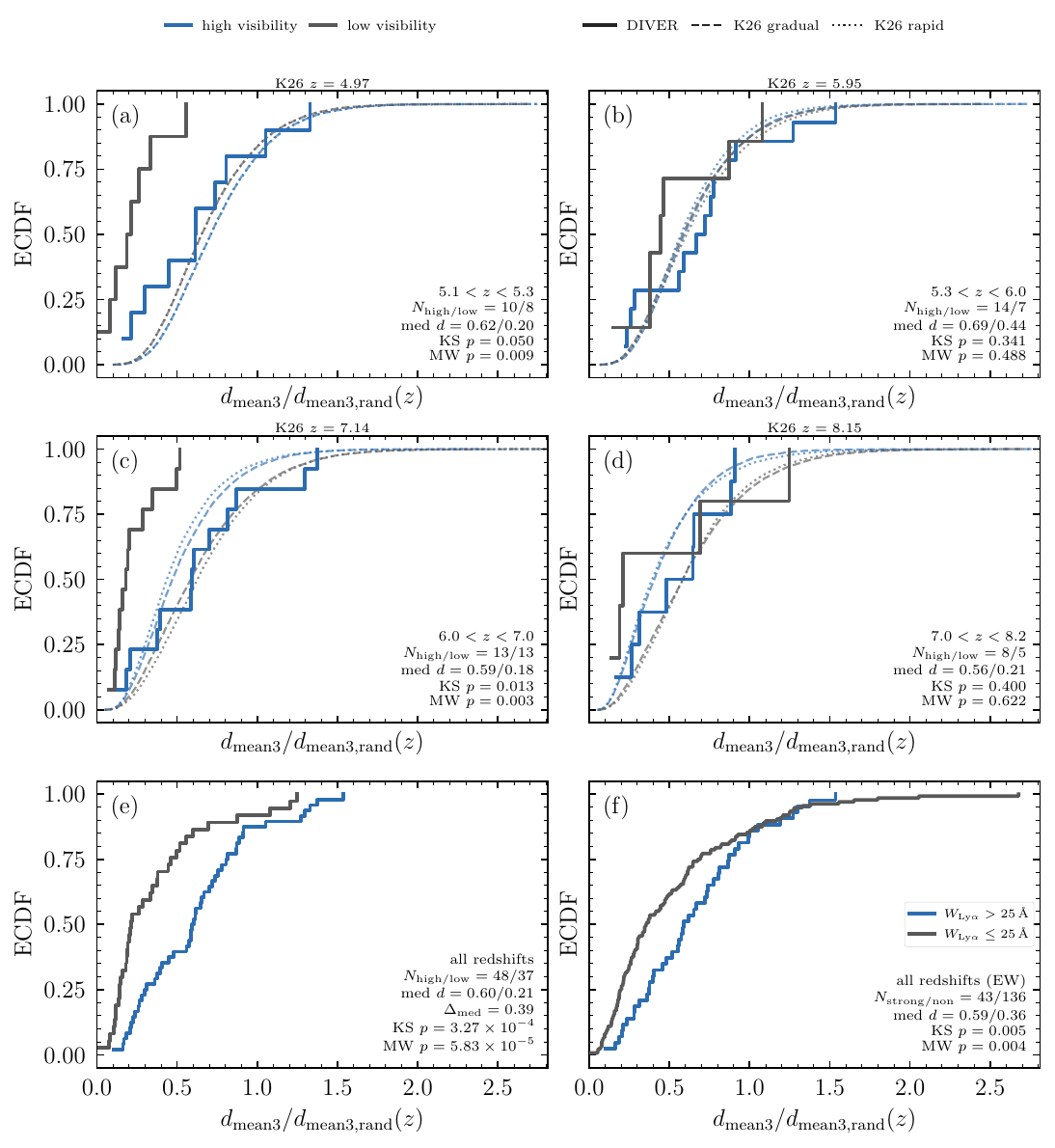}
    \caption{
    Ly$\alpha$ visibility as a function of normalized neighbor distance. \textbf{(a)}--\textbf{(d)} Cumulative distributions for the high- and low-visibility grating samples in four redshift intervals. High-visibility sources have detected $f_{\rm esc,Ly\alpha}^{\rm eff}>0.05$, while the low-visibility sample includes detections or constraining 3$\sigma$ upper limits at or below 0.05. Dashed and dotted curves show the K26 gradual and rapid models, with each simulated snapshot divided at its median $f_{\rm esc,IGM}$. \textbf{(e)} All-redshift comparison for the grating sample. \textbf{(f)} All-redshift comparison between 43 strong and 136 non-strong Ly$\alpha$ emitters, divided at $W_{\rm Ly\alpha}=25$~\AA. Annotations give the sample sizes, median normalized distances, and KS and Mann--Whitney probabilities. Values of $p<0.05$ indicate statistically significant differences between the distributions. Blue denotes higher Ly$\alpha$ visibility, and gray denotes lower visibility. Smaller normalized distances correspond to denser line-emitter environments.
    }
    \label{fig:normalized_neighbor_escape}
\end{figure*}

\section{Results}\label{sec:results}

We first examine the redshift dependence of Ly$\alpha$ visibility in GOODS-N and then test its relation to the local density field traced by WFSS line emitters.

\subsection{\texorpdfstring{Ly$\alpha$ visibility across redshift}{Lyalpha visibility across redshift}} \label{sec:results_z}

Figure~\ref{fig:lya_fraction_redshift} summarizes the redshift evolution of Ly$\alpha$ visibility in the DIVER sample. We use the fraction of galaxies with strong Ly$\alpha$ emission, defined as $W_{\rm Ly\alpha}>25$~\AA\ at EW ${\rm S/N}>3$, and the effective visibility $f_{\rm esc,Ly\alpha}^{\rm eff}$ defined in Section~\ref{sec:data_sed}. Panel (a) compares the strong-Ly$\alpha$ fraction with recent JWST/NIRSpec measurements from \citet{tang_jwstnirspec_2024,tang_ly_2024,napolitano_ly_2026}; see also \citet{nakane_ly_2024,napolitano_peering_2024,kageura_census_2025,jones_jades_2024,jones_jades_2025}. The K26 rapid and gradual reionization models are shown for comparison.

The DIVER measurements broadly follow the decline in the strong-Ly$\alpha$ fraction toward higher redshift seen in previous observations and the K26 models.
The point at $4.8\leq z<5.5$ falls below the smooth model trends, with 11 strong Ly$\alpha$ emitters among 82 grating sources. This fraction is similar to the UDS measurement from \citet{napolitano_ly_2026}, but lower than the measurement from \citet{tang_jwstnirspec_2024}, illustrating the substantial field-to-field variation at similar redshifts \citep[see also][]{ma_mammoth-subaru_2024}. This redshift interval contains the well-known $z\simeq5.2$ GOODS-N structure around HDF850.1 and an extended concentration of line-emitting galaxies \citep{walter_intense_2012,sun_jades_2024,helton_identification_2024,herard-demanche_mapping_2025}. In the WFSS catalog, the strongest peak in the number of line emitters occurs over $z\simeq5.1$--5.3. Figure~\ref{fig:overdensity_suppression} shows the structure in redshift space, together with the WFSS tracer counts and the strong-Ly$\alpha$ fraction. The shaded interval, $5.15<z<5.25$, marks the core of the overdensity, where the strong-Ly$\alpha$ fraction is low.

Figure~\ref{fig:lya_fraction_redshift}(b) shows the individual $f_{\rm esc,Ly\alpha}^{\rm eff}$ detections and 3$\sigma$ upper limits. The plotted sample contains 59 grating detections and 81 limits, together with 47 PRISM detections and 34 limits. Including the upper limits, the Kaplan--Meier median $\log_{10}f_{\rm esc,Ly\alpha}^{\rm eff}$ values for the grating sample are $-1.64$, $-1.26$, and $-1.28$ in the first three redshift intervals. The corresponding PRISM medians are $-0.59$, $-0.78$, and $-0.77$. Neither observing mode shows a monotonic change across these intervals because of the large source-to-source scatter. At $z\geq8.5$, the number of detections is too small to determine the median. In Section~\ref{sec:discussion_ceiling}, we show that the upper envelope of the visibility distribution provides a useful constraint on the reionization timeline.

Figure~\ref{fig:lya_properties_m1500} compares the DIVER Ly$\alpha$ measurements with the JADES LAE sample from \citet{jones_jades_2025} (see also \citealp{jones_jades_2024}). Panel (a) shows $W_{\rm Ly\alpha}$ as a function of $M_{1500,\rm obs}$. The DIVER grating sample extends to fainter UV luminosities than the JADES sample and spans a similar range of Ly$\alpha$ EW. Panel (b) shows that UV-fainter galaxies tend to have higher $f_{\rm esc,Ly\alpha}^{\rm eff}$, although with substantial scatter. Panel (c) shows the expected positive relation between $f_{\rm esc,Ly\alpha}^{\rm eff}$ and $W_{\rm Ly\alpha}$. Finally, panel (d) compares the Ly$\alpha$ EW distributions. The DIVER grating sample is broadly consistent with the JADES distribution, with a KS probability of $p=0.062$. The PRISM sample is weighted toward strong Ly$\alpha$ emitters and differs from the JADES distribution, with $p=3.3\times10^{-3}$, mainly because of its shallower Ly$\alpha$ sensitivity.

\subsection{\texorpdfstring{Dependence of Ly$\alpha$ visibility on environment}{Dependence of Lyalpha visibility on environment}} \label{sec:results_density}

Figure~\ref{fig:normalized_neighbor_escape} shows the main result of this work: galaxies with higher Ly$\alpha$ visibility lie farther from nearby WFSS line emitters. We quantify the environment using the normalized neighbor distance, $d_{\rm mean3}/d_{\rm mean3,rand}(z)$. Smaller values correspond to denser line-emitter environments, while larger values correspond to less dense environments. The typical projected separations are several cMpc, so the measurement traces the large-scale environment rather than close companions.

We first use the deeper grating sample and divide the galaxies at $f_{\rm esc,Ly\alpha}^{\rm eff}=0.05$. Detections above this threshold form the high-visibility sample. The low-visibility sample includes detections and constraining 3$\sigma$ upper limits at or below the threshold. Upper limits above 0.05 do not determine the classification and are omitted from this comparison. The resulting sample contains 48 high-visibility and 37 low-visibility galaxies, with another 47 galaxies left unclassified.

Over the full redshift range, the high-visibility galaxies have a median normalized neighbor distance of 0.598, compared with 0.211 for the low-visibility galaxies (Figure~\ref{fig:normalized_neighbor_escape}(e)). The difference is $\Delta_{\rm med}=0.386$. The two distributions give KS $p=3.27\times10^{-4}$ and Mann--Whitney $p=5.83\times10^{-5}$. Randomly reassigning the measured environments among the galaxies gives $p=2.00\times10^{-4}$, with only 0.02\% of the randomized samples producing a median difference at least as large as the observed one. Resampling the galaxies with replacement gives a 16th--84th percentile range of 0.286--0.424 for $\Delta_{\rm med}$, and every resampled value is positive. The same ordering is obtained when the visibility threshold is changed to 0.03 or 0.10.

The redshift-resolved comparisons in Figure~\ref{fig:normalized_neighbor_escape}(a)--(d) show the same ordering in all four intervals. The separation is strongest at $5.1<z<5.3$, where the high- and low-visibility samples have median normalized distances of 0.62 and 0.20, and at $6.0<z<7.0$, where the corresponding medians are 0.59 and 0.18. The Mann--Whitney probabilities are $p=0.009$ and $p=0.003$, respectively. The $5.3<z<6.0$ and $7.0<z<8.2$ intervals have the same direction of the difference, but the smaller samples do not show a significant separation. The K26 curves are shown for comparison. The simulated galaxies are divided at the median $f_{\rm esc,IGM}$ of each snapshot, while the observations use the fixed threshold $f_{\rm esc,Ly\alpha}^{\rm eff}=0.05$. We discuss the comparison with K26 further in Section~\ref{sec:discussion_sims}.

The Ly$\alpha$ EW comparison gives an independent view of the same result. The 43 galaxies with $W_{\rm Ly\alpha}>25$~\AA\ have a median normalized neighbor distance of 0.59, compared with 0.36 for the 136 non-strong emitters (Figure~\ref{fig:normalized_neighbor_escape}(f)). The two distributions give KS $p=0.005$ and Mann--Whitney $p=0.004$. Thus, the environmental difference does not depend on the H$\alpha$ normalization used to calculate $f_{\rm esc,Ly\alpha}^{\rm eff}$.

The result also remains when we retain all grating and PRISM detections and upper limits. Appendix~\ref{app:visibility_robustness} divides the galaxies into four equal-count intervals of normalized neighbor distance. The median $\log_{10}f_{\rm esc,Ly\alpha}^{\rm eff}$ increases from $-2.26$ in the densest interval to $-0.93$ in the least dense interval. A log-rank comparison gives $p=0.0044$, and the probability of obtaining an equally strong ordered trend by chance is $p=2.50\times10^{-4}$. The strong-Ly$\alpha$ fraction increases from 0.11 to 0.34 across the same intervals. The binary $f_{\rm esc,Ly\alpha}^{\rm eff}$ comparison, the EW comparison, and the analysis including upper limits all show lower Ly$\alpha$ visibility in denser regions of GOODS-N.

\begin{figure*}[!ht]
    \centering
    \includegraphics[width=\textwidth]{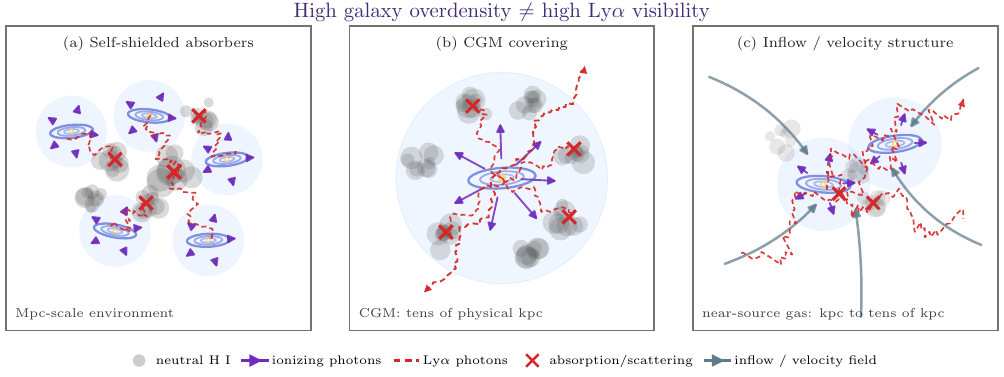}
    \caption{
    Possible mechanisms for low Ly$\alpha$ visibility in galaxy overdensities. \textbf{(a)} Self-shielded H\,\textsc{i} absorbers in the Mpc-scale environment. \textbf{(b)} Enhanced or patchy neutral-gas covering in the CGM. \textbf{(c)} Inflow and velocity structure in the near-source gas. Each mechanism can suppress or redirect Ly$\alpha$ photons despite the enhanced ionizing output of clustered galaxies. The sketches are not to scale, and the mechanisms are not mutually exclusive.
    }
    \label{fig:cartoon_topology}
\end{figure*}

\begin{figure*}[!ht]
    \centering
    \includegraphics[width=0.95\textwidth]{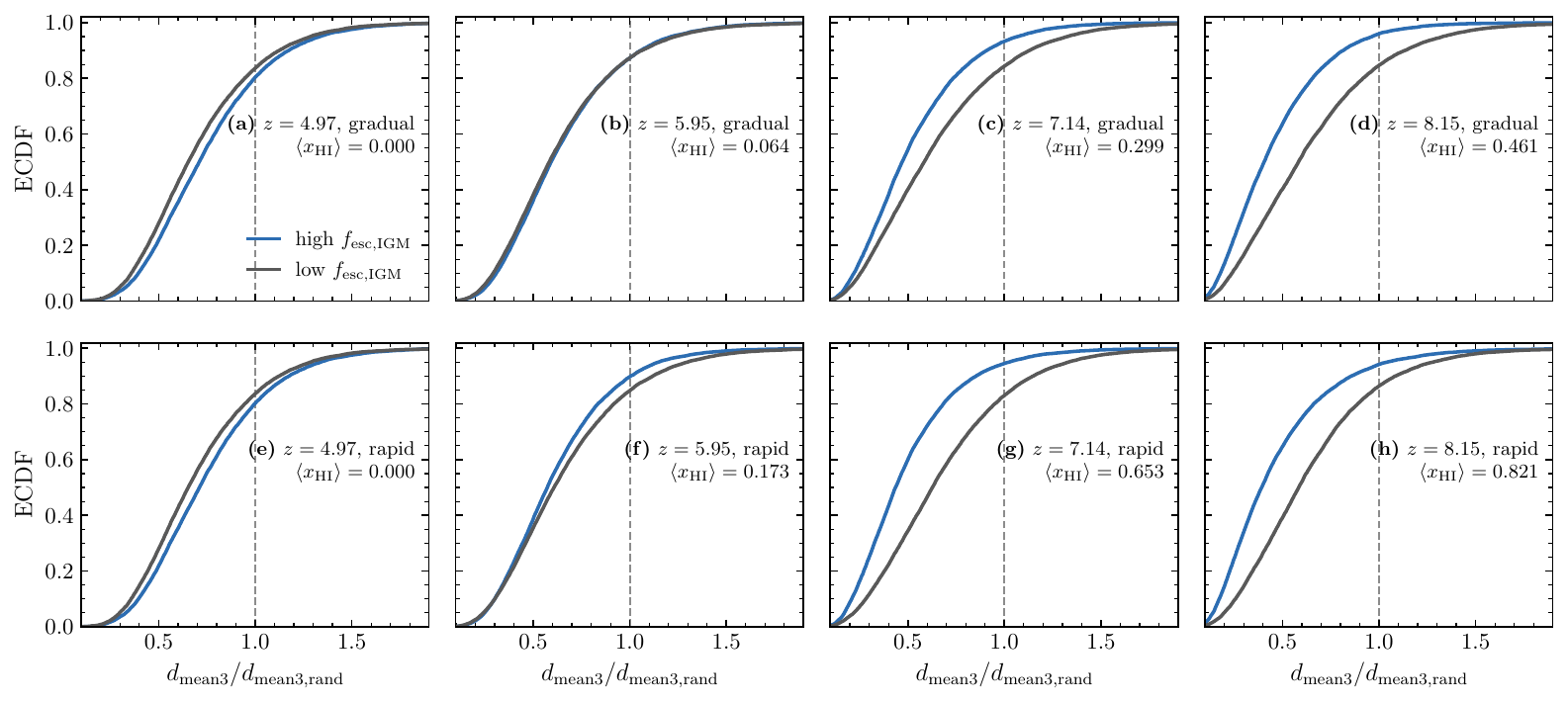}
    \caption{
    IGM transmission and environment in the K26 simulations. The top and bottom rows show the gradual and rapid reionization histories, respectively, with redshift increasing from left to right. Blue and gray curves show galaxies above and below the median $f_{\rm esc,IGM}$ in each snapshot. Each panel lists the redshift and volume-averaged neutral fraction, $\langle x_{\rm HI}\rangle$. The vertical dashed line marks a normalized neighbor distance of unity; smaller values correspond to denser environments. For the simulations, we use a three-dimensional nearest-neighbor statistic, whereas DIVER uses projected distances within $\pm10$~cMpc. The comparison is therefore qualitative.
    }
    \label{fig:keating_simulation_context}
\end{figure*}

\section{Implications for the Topology of Reionization} \label{sec:discussion}

The measurements above show that Ly$\alpha$ visibility may depend on both near-source radiative transfer and the large-scale galaxy environment. We first discuss the environmental trend in GOODS-N and several possible physical explanations, and then compare the measurements with reionization simulations. 

\subsection{\texorpdfstring{Density modulation of Ly$\alpha$ visibility}{Density modulation of Lyalpha visibility}} \label{sec:discussion_density}

The central result is that Ly$\alpha$ visibility is lower in denser line-emitter environments. The comparison divided at $f_{\rm esc,Ly\alpha}^{\rm eff}=0.05$, the EW-selected comparison, and the analysis retaining detections and upper limits all give the same ordering (Figure~\ref{fig:normalized_neighbor_escape}; Appendix~\ref{app:visibility_robustness}). The result also remains when the grating sample is divided at 0.03 or 0.10 and when we use alternative environment estimators.

This result does not conflict with the detection of strong LAEs in overdensities or inferred ionized regions \citep{saxena_jades_2023,saxena_jades_2024,witstok_inside_2024,witstok_jades_2025}. Those studies identify galaxies for which Ly$\alpha$ is transmitted, while DIVER measures the visibility distribution of all galaxies with usable spectral coverage, including non-detections. Ionized channels around clustered sources can produce bright LAEs even when many other galaxies in the same structure remain strongly attenuated \citep{hashemi_ly_2025,chen_impact_2026,li_reionization_2026}. The spatial distribution of detected LAEs provides complementary information about reionization morphology \citep[e.g.,][]{maitra_lyman_2025}.

The environmental ordering is broadly consistent with previous Ly$\alpha$ forest studies, which suggest that the most transmissive sightlines at $z\simeq5.7$ may lie away from LAE and [O\,\textsc{iii}] overdensities \citep{christenson_relationship_2023,zhu_galaxy_2026,jin_spectroscopic_2024}. As a separate check, Appendix~\ref{app:tau_checks} examines foreground Ly$\alpha$ forest transmission in the DIVER PRISM spectra \citep[following, e.g.,][]{meyer_probing_2025,hu_using_2026} and its relation to the WFSS density field. The $5.3<z<5.5$ interval shows a possible difference in forest transmission with environment, but the sample is small and the spectra are not deep enough for a firm conclusion. We therefore treat this analysis only as a supporting check.

The forest test should not be expected to reproduce the galaxy-visibility trend exactly, because Ly$\alpha$ forest transmission and galaxy Ly$\alpha$ visibility probe different opacity regimes. The forest traces resonant absorption by foreground intergalactic gas and depends on density, temperature, and the ionizing background. Galaxy Ly$\alpha$ visibility is more sensitive to radiative transfer near the source, including the ISM, CGM, local IGM, and the intrinsic line profile, as well as damping-wing absorption by substantially neutral gas farther away. Gas motions within a few comoving megaparsecs can also move photons closer to or blueward of resonance in the gas frame. The two observables can respond to the same large-scale structures while weighting the surrounding gas differently.

Ly$\alpha$ visibility is sensitive to gas over a broad range of scales. A galaxy overdensity can increase the local ionizing background and produce a large ionized region, but the observed line also depends on gas in the CGM and local IGM. Figure~\ref{fig:cartoon_topology} illustrates three possible ways to obtain low Ly$\alpha$ visibility in an overdensity. First, overdense regions may contain more self-shielded H\,\textsc{i} absorbers or high-column-density clumps, including structures that are not fully resolved in current reionization simulations. Such absorbers can produce broad or asymmetric suppression, although the line profile alone cannot distinguish among column density, covering fraction, velocity structure, source geometry, and large-scale IGM absorption. The damping-wing-like break in JADES-GS-z11-0, for example, can also be fit with a high-column-density proximate absorber \citep{hainline_searching_2024}. Second, galaxy-associated gas may have a larger or more patchy neutral-gas covering fraction. JWST/NIRSpec observations show that multiphase and kinematically disturbed gas is already present in galaxies during reionization \citep{zhu_early_2026,chen_spurs_2026,nakane_jwst_2026}. Third, coherent inflow or other near-source velocity structure can move Ly$\alpha$ photons closer to resonance in the gas frame, increasing resonant absorption even when the surrounding IGM is highly ionized \citep{keating_jwst_2024,cain_chasing_2025}. These mechanisms are not mutually exclusive and help explain why galaxy overdensity and Ly$\alpha$ visibility do not always track one another. Interactions can also increase Ly$\alpha$ visibility: merger-driven bursts may open low-column-density channels and enhance Ly$\alpha$ escape \citep{witten_deciphering_2024}.

We also tested whether the environmental trend could be explained by differences in galaxy properties. Appendix~\ref{app:visibility_robustness} shows a marginal difference in the dust distributions, although it is not supported consistently by both statistical tests. The stellar-mass and H$\alpha$/UV ionizing-production comparisons show no significant differences. None of these quantities shows an environmental separation as strong or as consistently supported as the Ly$\alpha$ measurements. However, we caution that these tests do not rule out a contribution from galaxy properties. Stellar mass, star-formation rate, metallicity, dust geometry, star-formation history, and the intrinsic Ly$\alpha$ profile can all affect the observed emission and may themselves depend on environment. Larger samples with uniform measurements of these quantities will be needed to separate their contributions. Strong Ly$\alpha$ emitters also have lower median SED-inferred dust attenuation than non-strong emitters. This is consistent with a contribution from dust, but does not show that dust drives the environmental trend.

\subsection{Comparison with simulations} \label{sec:discussion_sims}

We compare the DIVER measurements with the post-processed K26 reionization simulations. K26 uses the Sherwood-Relics cosmological hydrodynamical simulations \citep{puchwein_sherwood-relics_2023}, run with a modified version of \textsc{p-gadget-3} based on \textsc{gadget-2} \citep{springel_cosmological_2005}. We use the $160~h^{-1}$~cMpc box containing $2\times2048^3$ particles, with dark-matter and gas particle masses of $M_{\rm DM}=5.07\times10^7~M_\odot$ and $M_{\rm gas}=9.41\times10^6~M_\odot$. K26 post-processes the simulations with the ATON radiative-transfer code \citep{aubert_radiative_2008} using gradual and rapid reionization histories. The Ly$\alpha$ transmission is calculated through the simulated density, temperature, ionization, and velocity fields. The spectra are generated in the halo rest frame, so inflows and other local motions can alter the transmission.

Figure~\ref{fig:keating_simulation_context} shows the relation between Ly$\alpha$ transmission and environment in K26. Halos represent both the target galaxies and the density tracers, without assigning H$\alpha$ or [O\,\textsc{iii}] luminosities. For each target halo, we calculate the three-dimensional mean distance to the nearest three tracer halos and normalize it by the corresponding value from 500 random positions. We note that this differs from the projected DIVER statistic measured within $\pm10$~cMpc. We do not model line luminosities, duty cycles, or observational detection limits, so we compare only how Ly$\alpha$ transmission varies with environment.

For Figure~\ref{fig:keating_simulation_context}, we divide the galaxies in each K26 snapshot at the median modeled $f_{\rm esc,IGM}$. We introduce this division to separate galaxies with relatively higher and lower IGM transmission. The simulations show that the sign of the environment--transmission relation changes with the dominant source of opacity. At volume-averaged neutral fractions $\langle x_{\rm HI}\rangle\gtrsim0.1$, galaxies near other tracers tend to lie in larger ionized regions and stronger local ionizing backgrounds. Galaxies with higher modeled transmission therefore lie closer to nearby tracers, as expected for inside-out reionization.

At lower neutral fractions, density-dependent resonant opacity in ionized gas becomes more important. For gas in photoionization equilibrium,
\begin{equation}
    \tau_{\rm IGM}\propto \Gamma_{\rm HI}^{-1}T^{-0.7}\Delta^2,
\end{equation}
where $\Gamma_{\rm HI}$ is the hydrogen photoionization rate, $T$ is the gas temperature, and $\Delta$ is the gas overdensity. Dense gas around galaxies and filaments can then reduce Ly$\alpha$ transmission, especially when peculiar velocities or inflows move the photons close to resonance in the gas frame. An enhanced ionizing background can partly offset this density dependence, so the relation depends on the relative variations in $\Gamma_{\rm HI}$, temperature, and density. This behavior is consistent with the EIGER galaxy--Ly$\alpha$ forest cross-correlation measurements \citep{kashino_eiger_2025}. It also shows why galaxy Ly$\alpha$ visibility and Ly$\alpha$ forest transmission need not follow the same environmental trend: they weight near-source radiative transfer and foreground resonant opacity differently. This density-dependent resonant opacity is distinct from the damping-wing attenuation considered in Section~\ref{sec:discussion_ceiling}.

For a relative comparison between the observations and simulations, we divide each observational redshift interval at its Kaplan--Meier median and each K26 snapshot at its median modeled $f_{\rm esc,IGM}$. The DIVER comparison gives $\Delta_{\rm med}=0.51$, 0.36, 0.42, and 0.44 from the lowest to the highest redshift interval. The first value is descriptive because the low-visibility sample contains only four galaxies. The nearest K26 snapshots give gradual/rapid values of $+0.057/+0.058$ at $z=4.97$, $+0.009/-0.034$ at $z=5.95$, $-0.115/-0.191$ at $z=7.14$, and $-0.178/-0.192$ at $z=8.15$. Appendix~\ref{app:simulation_comparisons} presents this comparison in full.

The lowest-redshift DIVER interval has the same positive sign as K26 but a larger separation. At $5.3<z<6.0$, DIVER remains positive while K26 is close to zero. At $z>6$, DIVER continues to show lower visibility in denser environments, while K26 predicts the opposite ordering. The observational separation is also larger than the K26 result in every interval, although the amplitudes cannot be compared directly. Part of the difference may arise because the DIVER measurements include ISM, CGM, local-IGM, aperture, galaxy-property, and H$\alpha$-normalization effects. The observations and simulations also use different tracer selections and environment statistics. The sign difference at $z\gtrsim6$ suggests that the global neutral fraction alone does not determine the observed environmental trend. CGM and local-IGM opacity, unresolved high-column-density absorbers, source selection, and measurement effects may all contribute.

We also tested whether the K26 relation depends strongly on tracer-halo mass. We repeated the calculation using the full selected halo sample, the highest-mass quartile, and the lowest-mass quartile as tracers, with equal numbers of halos in the two quartile samples. For the gradual model at $z=5.95$, the separation remains consistent with zero for all three selections, with $|\Delta_{\rm med}|<0.01$. The positive relation at lower redshift and the negative relation at higher redshift are otherwise qualitatively unchanged.

Appendix~\ref{app:simulation_comparisons} presents an independent qualitative comparison with the FlexRT reionization histories \citep{cain_flexrt_2024,cain_chasing_2025}. These models span late, early, and very early starts to reionization, with all three histories ending near $z\sim5$. The available products do not provide a direct $f_{\rm esc,IGM}$ value. We therefore calculate the damping-wing opacity from the forward-modeled spectra and use it to rank the sources by relative Ly$\alpha$ visibility. At higher neutral fractions, sources with higher relative visibility tend to lie closer to nearby tracers. At lower neutral fractions, the separation weakens or reverses as density-dependent opacity becomes more important. The FlexRT comparison therefore supports the same broad interpretation as K26.

\begin{figure*}[!ht]
    \centering
    \includegraphics[width=\textwidth]{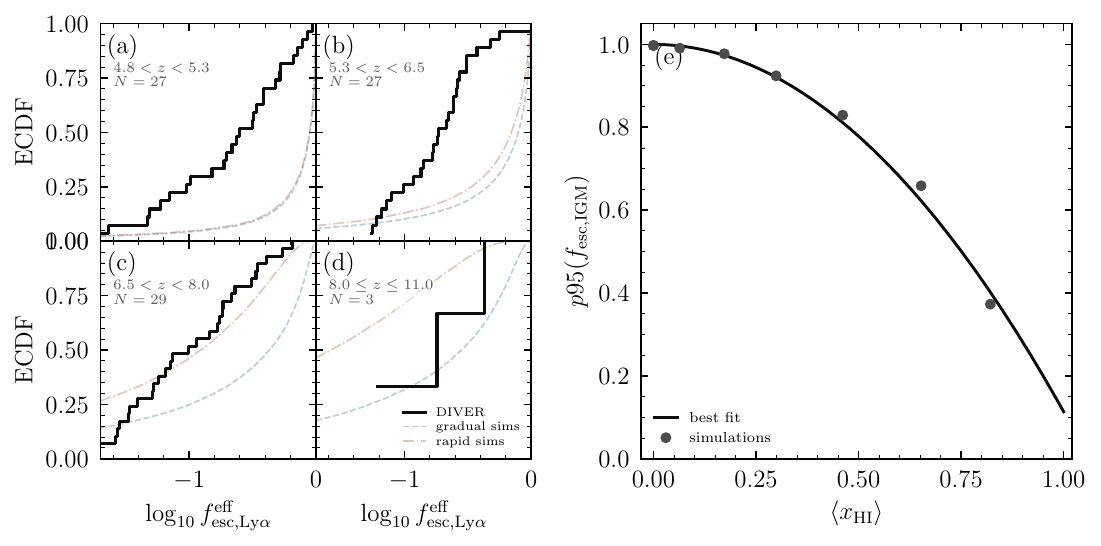}
    \caption{
    Upper envelope of Ly$\alpha$ visibility and its calibration with K26. \textbf{(a)}--\textbf{(d)} Cumulative distributions of $f_{\rm esc,Ly\alpha}^{\rm eff}$ for the DIVER Ly$\alpha$ detections in four redshift intervals. Black curves show DIVER, while dashed and dot-dashed curves show the K26 gradual and rapid reionization histories, respectively. \textbf{(e)} The 95th percentile of $f_{\rm esc,IGM}$ as a function of the volume-averaged neutral fraction. Gray points show the K26 snapshots, and the black curve shows the best-fit relation $p95(f_{\rm esc,IGM})=\max[0,1-A\langle x_{\rm HI}\rangle^2]$, with $A=0.886$. Panels~(a)--(d) use broader redshift intervals for display; the neutral-fraction ceilings are calculated in the finer intervals listed in Table~\ref{tab:upper_envelope}.
    }
    \label{fig:neutral_fraction_constraint}
\end{figure*}

\begin{figure*}[!ht]
    \centering
    \includegraphics[width=0.88\textwidth]{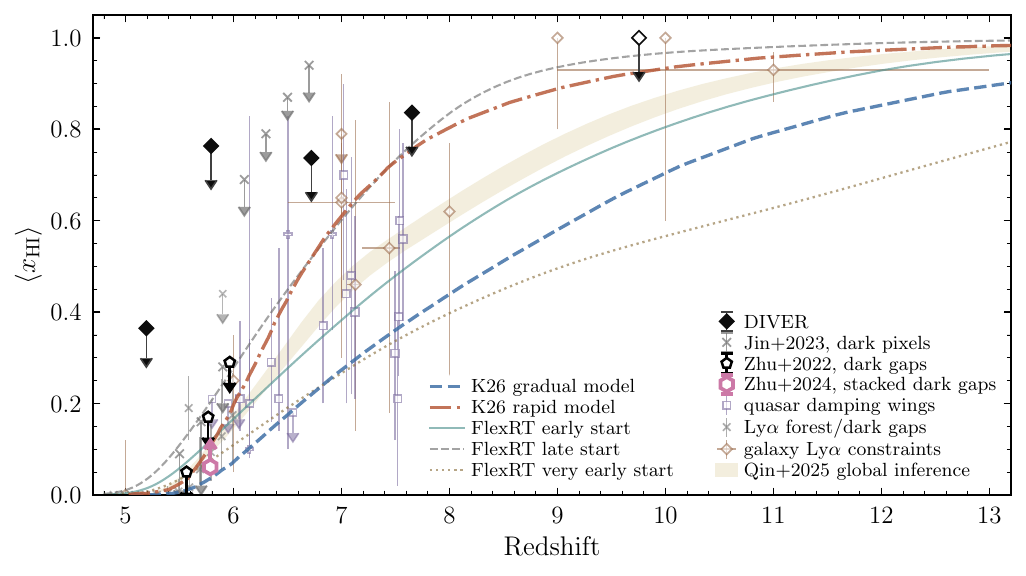}
    \caption{
    DIVER neutral-fraction ceilings compared with published constraints and reference reionization histories. Filled black diamonds show the DIVER ceilings inferred from the upper envelope of $f_{\rm esc,Ly\alpha}^{\rm eff}$, with downward arrows indicating the upper-limit direction. The open black diamond marks the $z\simeq9.8$ bin, which contains one Ly$\alpha$ detection. Literature constraints are grouped by method \citep{spina_damping_2024,greig_igm_2024,durovcikova_chronicling_2024,umeda_jwst_2023,jin_nearly_2023,zhu_long_2022,zhu_damping_2024,gaikwad_measuring_2023,greig_igm_2022,yang_poniuaena_2020,wang_significantly_2020,greig_constraints_2019,davies_determining_2018,banados_800-million-solar-mass_2018,greig_are_2017,jones_jades_2025,nakane_ly_2024,umeda_probing_2026}. The shaded band shows the global inference from \citet{qin_percent-level_2025}. The curves show the K26 gradual and rapid histories and the FlexRT early-start, late-start, and very-early-start histories.
    }
    \label{fig:neutral_fraction_comparison}
\end{figure*}

\section{Neutral-fraction ceilings} \label{sec:discussion_ceiling}

We use the upper tail of the $f_{\rm esc,Ly\alpha}^{\rm eff}$ distribution to place one-sided limits on the neutral fraction. If the most Ly$\alpha$-visible galaxies preferentially sample the clearest sightlines, increasing IGM opacity should lower this upper envelope. Assigning all of the suppression to the IGM gives the largest neutral fraction consistent with the observed envelope. Additional attenuation in the ISM, CGM, or aperture generally lowers the observed visibility and makes this interpretation conservative. This is similar in spirit to dark-pixel and dark-gap methods, which use one-sided transmission statistics without assigning every dark region to a unique origin \citep[e.g.,][]{mcgreer_model-independent_2015,zhu_long_2022,zhu_damping_2024,jin_nearly_2023,davies_updated_2025}. Unlike the dark-pixel bound, however, the conversion developed here requires calibration with reionization simulations.

For the ceiling calibration, we focus on damping-wing absorption by neutral gas outside the local ionized region \citep[e.g.,][]{miralda-escude_searching_1998,mesinger_ly_2008,keating_origin_2024}. Schematically,
\begin{equation}
\tau_{\rm DW}(\Delta v)
\simeq
\int_{R_b}^{\infty}
n_{\rm HI}(r)\,
\sigma_{\alpha}\!\left[\Delta v+H(z)r\right]\,
dr,
\end{equation}
where $r$ and $R_b$ are proper distances and $R_b$ is the distance to the first neutral structure along the sightline. In the damping wing, $\sigma_{\alpha}$ scales approximately as the inverse square of the velocity offset. Approximating the optical-depth-weighted neutral density beyond $R_b$ as proportional to $\langle x_{\rm HI}\rangle$ gives
\begin{equation}
\tau_{\rm DW}
\propto
\langle x_{\rm HI}\rangle
\int_{R_b}^{\infty}
\frac{dr}{[\Delta v+H(z)r]^2}
\propto
\frac{\langle x_{\rm HI}\rangle}
{\Delta v+H(z)R_b}.
\label{eq:damping_wing_rb}
\end{equation}
For $H(z)R_b\gg\Delta v$, this reduces to
$\tau_{\rm DW}\propto\langle x_{\rm HI}\rangle/R_b$. In a patchy IGM, $R_b$ varies substantially among sightlines because the neutral gas is distributed in discrete structures rather than as a uniform screen.

We use a simple model to motivate how $R_b$ may scale for the most transmissive sightlines. Let $\lambda_{\rm n}$ be the mean number of neutral structures intercepted per unit proper path length. If the encounters are approximately Poisson distributed,
\begin{equation}
P(R_b>L)=\exp(-\lambda_{\rm n}L),
\end{equation}
and the $q$th percentile of the distance distribution is
\begin{equation}
R_{b,q}
=
-\frac{\ln(1-q)}{\lambda_{\rm n}}.
\end{equation}
If the characteristic sizes and geometry of the neutral structures vary slowly, their line-of-sight incidence scales approximately as
$\lambda_{\rm n}\propto\langle x_{\rm HI}\rangle$. At fixed $q$, this gives
$R_{b,q}\propto\langle x_{\rm HI}\rangle^{-1}$. Combining this relation with Equation~\ref{eq:damping_wing_rb} gives, for the high-transmission tail,
\begin{equation}
\tau_{\rm DW}^{(q)}
\propto
\frac{\langle x_{\rm HI}\rangle}{R_{b,q}}
\propto
\langle x_{\rm HI}\rangle^2.
\end{equation}
The corresponding transmission scales as
\begin{equation}
T_{\rm IGM}^{(q)}
\simeq
\exp\!\left[-C\langle x_{\rm HI}\rangle^2\right]
\simeq
1-C\langle x_{\rm HI}\rangle^2
\end{equation}
at low optical depth.

This argument motivates the quadratic dependence, although we caution that it may not be a general law of reionization topology. Evolution in the neutral-island size distribution, interface area, source bias, Ly$\alpha$ velocity offsets, or sightline selection can change the scaling. We therefore calibrate the upper-tail relation directly from the simulations rather than attempting to fix its normalization from this schematic model.

The sample maximum is sensitive to individual outliers and sample size, so we use the 95th percentile as the upper-envelope statistic. For each redshift interval, we define
\begin{equation}
\mathcal{T}_{95}(z)
\equiv
p95\!\left(f_{\rm esc,Ly\alpha}^{\rm eff}\mid z\right).
\end{equation}
We calculate $\mathcal{T}_{95}$ from 86 galaxies with Ly$\alpha$ detections, using the grating measurement when available and the PRISM measurement otherwise. Of these, 84 have the environment measurements used in Section~\ref{sec:results_density}. The non-detections do not provide measured values for the upper tail. We estimate the sampling uncertainty by bootstrapping the galaxies within each redshift interval.

We fit the K26 snapshots with
\begin{equation}
p95(f_{\rm esc,IGM})
=
\max\!\left[
0,\,
1-A\langle x_{\rm HI}\rangle^2
\right].
\label{eq:upper_percentile_xhi_calibration}
\end{equation}
The combined gradual and rapid histories give $A=0.886$. Figure~\ref{fig:neutral_fraction_constraint}(a)--(d) compares the observed $f_{\rm esc,Ly\alpha}^{\rm eff}$ distributions with the simulated $f_{\rm esc,IGM}$ distributions, while panel (e) shows the percentile calibration. The shape of the simulated relation is well described by the quadratic form motivated above.

The ceiling follows from the one-sided nature of the comparison. If attenuation outside the IGM can only reduce the observed upper envelope, then
\begin{equation}
\mathcal{T}_{95}^{\rm obs}
\leq
p95\!\left(
f_{\rm esc,IGM}\mid
\langle x_{\rm HI}\rangle_{\rm true}
\right).
\end{equation}
Because Equation~\ref{eq:upper_percentile_xhi_calibration} decreases with neutral fraction, inverting the observed value gives
\begin{equation}
\langle x_{\rm HI}\rangle_{\rm true}
\leq
\langle x_{\rm HI}\rangle_{\rm max}
=
\min\!\left\{
1,\,
\left[
\frac{\max(0,1-\mathcal{T}_{95})}{A}
\right]^{1/2}
\right\}.
\label{eq:xhi_ceiling}
\end{equation}
The outer limit enforces $\langle x_{\rm HI}\rangle\leq1$, while the inner maximum keeps the inversion defined when the effective visibility exceeds unity.

The environmental comparison and the ceiling calibration use different aspects of K26. The former compares transmission with neighbor distance, while the latter uses only how the 95th percentile of transmission changes with the global neutral fraction. The disagreement in the environment relation therefore does not enter Equation~\ref{eq:xhi_ceiling} directly, although unresolved physics that changes the simulated upper envelope remains a systematic uncertainty.

One caveat is that additional ISM, CGM, or aperture attenuation lowers $\mathcal{T}_{95}$ and makes the inferred ceiling less restrictive. An underestimated H$\alpha$ denominator raises $f_{\rm esc,Ly\alpha}^{\rm eff}$ and can instead make the ceiling too restrictive. The grating--PRISM checks in Appendix~\ref{app:data_validation} limit large global normalization offsets but do not remove these systematic uncertainties.

\begin{deluxetable}{ccc}
\tablecaption{Ly$\alpha$ visibility upper envelope and neutral-fraction ceilings\label{tab:upper_envelope}}
\tablehead{
\colhead{Redshift range} &
\colhead{$p95(f_{\rm esc,Ly\alpha}^{\rm eff})$} &
\colhead{$\langle x_{\rm HI}\rangle_{\rm max}$}
}
\startdata
$4.8<z<5.5$   & $0.882^{+0.463}_{-0.146}$ & $0.36^{+0.18}_{-0.36}$ \\
$5.5<z<6.5$   & $0.483^{+0.083}_{-0.165}$ & $0.76^{+0.11}_{-0.06}$ \\
$6.5<z<7.5$   & $0.518^{+0.110}_{-0.172}$ & $0.74^{+0.12}_{-0.09}$ \\
$7.5<z<8.5$   & $0.380^{+0.050}_{-0.154}$ & $0.84^{+0.10}_{-0.03}$ \\
$8.5<z<9.5$   & \nodata                    & \nodata \\
$9.5<z<11.0$  & $0.0615\ (N=1)$           & $1.0\ (N=1)$ \\
\enddata
\tablecomments{The uncertainties give the 16th--84th percentile range from bootstrap resampling of the galaxies and are propagated through Equation~\ref{eq:xhi_ceiling}. Neutral-fraction values are limited to the physical interval $0\leq\langle x_{\rm HI}\rangle\leq1$. The $8.5<z<9.5$ interval contains no $f_{\rm esc,Ly\alpha}^{\rm eff}$ detection. The $9.5<z<11.0$ interval contains one detection, so both values are set by that object and no sampling uncertainty is reported.}
\end{deluxetable}

Table~\ref{tab:upper_envelope} gives the observed upper-envelope measurements. Equation~\ref{eq:xhi_ceiling} gives $\langle x_{\rm HI}\rangle_{\rm max}=0.36$, 0.76, 0.74, 0.84, and 1.0 at $z\simeq5.2$, 5.8, 6.7, 7.7, and 9.8, respectively. The bootstrap uncertainties on $\mathcal{T}_{95}$ are propagated through the same equation. The $z\simeq9.8$ value is based on one galaxy and provides no constraint beyond the physical limit $\langle x_{\rm HI}\rangle\leq1$.

We also test the effect of the quoted measurement uncertainties by perturbing each source measurement before recalculating $\mathcal{T}_{95}$. This gives ceilings of 0.25, 0.71, 0.64, and 0.69 in the first four populated intervals. Adding measurement scatter tends to raise the upper tail of the observed distribution, which lowers the inferred neutral-fraction ceiling. We treat these values only as a sensitivity test because perturbing the measured values broadens the distribution rather than recovering its intrinsic upper envelope. We therefore retain the source-bootstrap ceilings, which are weaker and hence more conservative.

Figure~\ref{fig:neutral_fraction_comparison} compares the DIVER ceilings with published volume-averaged neutral-fraction constraints and reference reionization histories. The $7.5<z<8.5$ interval gives $\langle x_{\rm HI}\rangle_{\rm max}=0.84$. Under the limiting IGM-attenuation interpretation, reionization was already underway by $z\simeq8$, and an almost completely neutral IGM is disfavored at this epoch. This is consistent with evidence for Ly$\alpha$ transmission at still earlier times, including the reported $z\simeq13$ Ly$\alpha$ emitter \citep{witstok_witnessing_2025,cohon_long_2025}. Favorable reionization morphology and intrinsic Ly$\alpha$ velocity offsets can increase the visibility of such sources \citep{qin_reionization_2025}. The present ceilings do not distinguish between rapid and gradual reionization histories. Larger samples will be needed to tighten these limits and distinguish among reionization histories.

\section{Conclusions} \label{sec:summary}

We used DIVER JWST/NIRSpec observations in GOODS-N to study the environmental dependence of galaxy Ly$\alpha$ visibility over $4.8<z<11$ and to constrain the reionization timeline. After applying the spectral-quality and wavelength-coverage cuts, the parent sample contains 262 galaxies. The EW-classification sample contains 250 galaxies; 179 have environment measurements, and 177 enter the continuous visibility--environment analysis. We combine the DIVER Ly$\alpha$ measurements with CONGRESS and FRESCO H$\alpha$ and [O\,\textsc{iii}] emitters to trace the surrounding line-emitter density field. Our main conclusions are:

\begin{enumerate}
    \item The strong-Ly$\alpha$ fraction broadly declines toward higher redshift. Its lowest nonzero value occurs at $4.8\leq z<5.5$, which contains the strongest WFSS line-emitter overdensity in GOODS-N at $z\simeq5.2$. Together with differing measurements in other fields at similar redshifts, this supports substantial field-to-field variation in Ly$\alpha$ visibility.

    \item Galaxies with higher Ly$\alpha$ visibility tend to lie farther from nearby WFSS line emitters. In the grating sample, the median normalized neighbor distances are 0.60 and 0.21 for the high- and low-visibility populations, respectively. The EW comparison and the analysis retaining the grating and PRISM upper limits give the same environmental trend.

    \item Dust attenuation, stellar mass, and the H$\alpha$/UV ionizing-production proxy do not show an environmental separation as strong or as consistently supported as Ly$\alpha$. Other galaxy properties, including star-formation rate, metallicity, dust geometry, star-formation history, and the intrinsic Ly$\alpha$ profile, may still contribute. Larger samples with more uniform measurements will be needed to quantify these effects.

    \item We compare the DIVER measurements with the K26 reionization simulations and find that the sign of the environment--transmission relation can change as the dominant opacity source evolves. DIVER shows lower Ly$\alpha$ visibility in denser environments across all four redshift intervals, including at $z>6$, where K26 predicts the opposite trend. The global neutral fraction alone therefore does not determine the observed environmental relation.

    \item We introduce an upper-envelope method for constraining the neutral fraction by calibrating the 95th percentile of $f_{\rm esc,Ly\alpha}^{\rm eff}$ with K26. Under the limiting assumption that the decline in this envelope is caused by IGM attenuation, we obtain neutral-fraction ceilings of $\langle x_{\rm HI}\rangle_{\rm max}=0.36$, 0.76, 0.74, 0.84, and 1.0 at $z\simeq5.2$, 5.8, 6.7, 7.7, and 9.8, respectively. The $z\simeq8$ ceiling disfavors an almost completely neutral IGM at this epoch.
\end{enumerate}

A Ly$\alpha$ redshift trend measured in one field can combine global reionization evolution with the density structure, CGM covering fraction, and absorber population of that field. Galaxy overdensities can therefore show either enhanced or suppressed Ly$\alpha$ visibility, depending on whether ionized-region growth or local opacity dominates. Neutral-fraction analyses based on Ly$\alpha$ emitter fractions or equivalent-width distributions should account for the sampled environment and tracer selection. Multi-field JWST surveys combining Ly$\alpha$ spectroscopy with H$\alpha$ and [O\,\textsc{iii}] density maps will help distinguish global reionization evolution from field-dependent radiative transfer.

\begin{acknowledgements}
Y.Z.\ acknowledges support from the NIRCam Science Team contract to the University of Arizona, NAS5-02105. Y.Z.\ is also supported by JWST program \#6434.
G.D.B.\ is supported by JWST Program \#4092.
L.K.\ acknowledges the support of a Royal Society University Research Fellowship (grant number URF$\backslash$R1$\backslash$251793).
C.C.\ acknowledges support from the Beus Center for Cosmic Foundations.
A.J.B.\ acknowledges funding from the ``FirstGalaxies'' Advanced Grant from the European Research Council under the European Union's Horizon 2020 research and innovation programme (grant agreement No.\ 789056).
J.M.H. acknowledges support from the Evolving Universe Fellowship, which is made possible by a generous donation from Dr. Keiko Miwa Ross; J.M.H. also acknowledges support from JWST Program \#8544.
R.M.\ and F.D.E.\ acknowledge support by the Science and Technology Facilities Council (STFC), by the ERC through Advanced Grant 695671 ``QUENCH'', and by the UKRI Frontier Research grant RISEandFALL.
R.M.\ also acknowledges funding from a research professorship from the Royal Society.
Support for programs \#4092, \#6434, and \#8544 was provided by NASA through a grant from the Space Telescope Science Institute, which is operated by the Association of Universities for Research in Astronomy, Inc., under NASA contract NAS 5-03127.
This manuscript benefited from language editing with ChatGPT \citep{openai_chatgpt_2026}. The authors are solely responsible for the scientific content. 

This work is based on observations made with the NASA/ESA/CSA James Webb Space Telescope. The data were obtained from the Mikulski Archive for Space Telescopes at the Space Telescope Science Institute, which is operated by the Association of Universities for Research in Astronomy, Inc., under NASA contract NAS 5-03127 for JWST. These observations are associated with program \#8018.
The authors acknowledge the FRESCO (PI: P.~Oesch) team for developing their observing program with a zero-exclusive-access period.

Some results supporting this work were obtained using computational resources distributed under NSF ACCESS allocations TG-PHY230158 and TG-PHY240332.  

We respectfully acknowledge the University of Arizona is on the land and territories of Indigenous peoples. Today, Arizona is home to 22 federally recognized tribes, with Tucson being home to the O'odham and the Yaqui. The university strives to build sustainable relationships with sovereign Native Nations and Indigenous communities through education offerings, partnerships, and community service.

\end{acknowledgements}

\vspace{5mm}
\facilities{JWST; MAST}

\software{
{\tt astropy} \citep{astropy_collaboration_astropy_2013,astropy_collaboration_astropy_2018,astropy_collaboration_astropy_2022},
{\tt JWST Calibration Pipeline} \citep{bushouse_jwst_2025},
{\tt scipy} \citep{virtanen_scipy_2020}
}

\appendix

\section{Data Products and Measurement Validation}
\label{app:data_validation}

Table~\ref{tab:source_measurement_summary} presents the source catalog used in this work. It contains 261 unique galaxies with Ly$\alpha$ or $f_{\rm esc,Ly\alpha}^{\rm eff}$ measurements or limits; an environment measurement is not required for inclusion. The table lists source positions and redshifts, G140M and PRISM Ly$\alpha$ measurements, PRISM- and SED-based H$\alpha$ measurements, the WFSS H$\alpha$ luminosity used for validation, the effective Ly$\alpha$ visibility, and the available environment measurements.

\input{source_measurement_summary.tex}

\begin{figure*}[!ht]
    \centering
    \includegraphics[width=\textwidth]{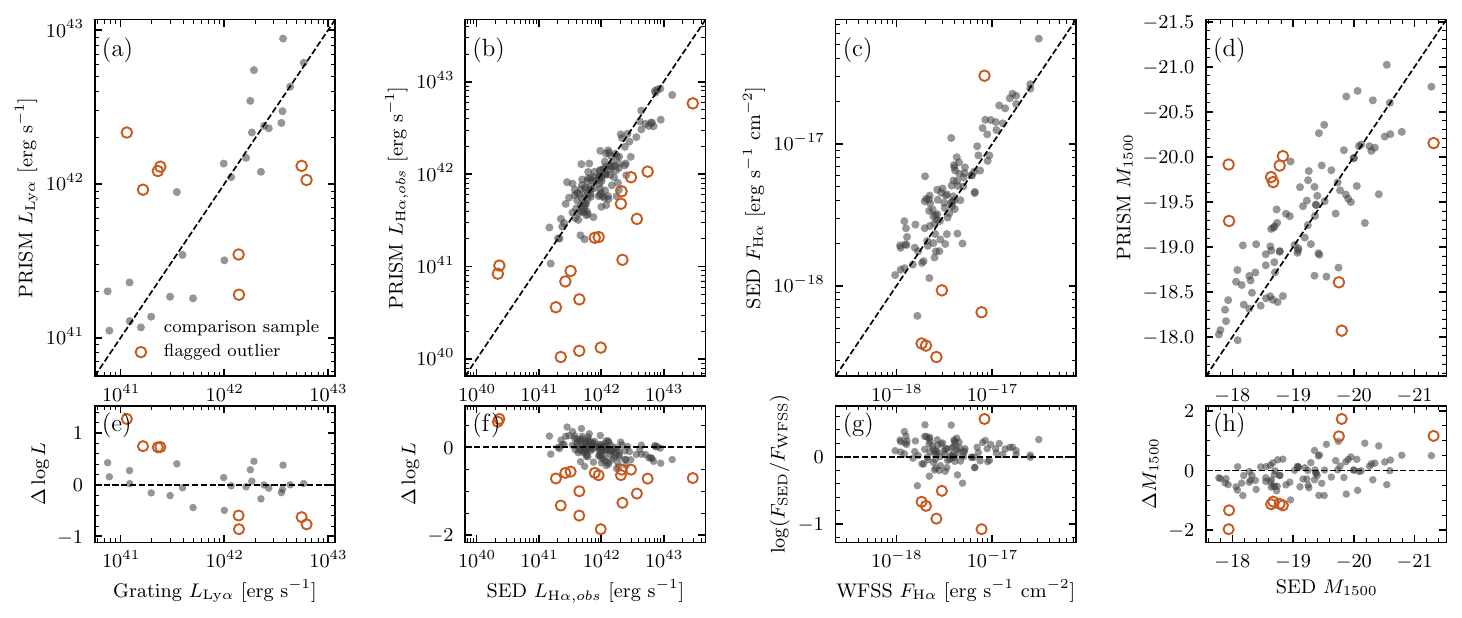}
    \caption{
    Comparison of the measurements used in the Ly$\alpha$ visibility analysis. \textbf{(a)} G140M and PRISM Ly$\alpha$ luminosities. \textbf{(b)} SED-based and PRISM H$\alpha$ luminosities. \textbf{(c)} SED-based and WFSS H$\alpha$ fluxes. \textbf{(d)} SED-based and PRISM $M_{1500}$ measurements. \textbf{(e)}--\textbf{(h)} Differences from equality for panels \textbf{(a)}--\textbf{(d)}. Open circles mark measurements differing by more than 0.5 dex in luminosity or flux, or by more than 1 mag in $M_{1500}$. Dashed lines show equality.
    }
    \label{fig:measurement_consistency}
\end{figure*}

We next compare the G140M, PRISM, SED, and WFSS measurements used in the Ly$\alpha$ visibility analysis. For the G140M sample, Ly$\alpha$ is measured from the grating spectrum, while the UV continuum and observed H$\alpha$ luminosity are inferred from the photometric SED fit. For the PRISM sample, all three quantities are measured from the PRISM spectrum. Figure~\ref{fig:measurement_consistency} shows the comparisons for sources with overlapping measurements.

The PRISM--SED H$\alpha$ comparison contains 147 sources with positive measurements. The median $\log_{10}(F_{\rm SED}/F_{\rm PRISM})$ is $+0.064$ dex, with a scatter of 0.223 dex. The WFSS--SED comparison contains 110 sources matched through the shared JADES v0.9.2 identifier. We require $\texttt{flag\_DR}=1$, H$\alpha$ ${\rm S/N}>3$, positive SED H$\alpha$, and $|z_{\rm SED}-z_{\rm WFSS}|\leq0.03$. This sample gives a median $\log_{10}(F_{\rm SED}/F_{\rm WFSS})=+0.072$ dex and a scatter of 0.170 dex. Six sources differ by more than 0.5 dex and are marked in Figure~\ref{fig:measurement_consistency}.

The comparisons show no large overall offset among the G140M, PRISM, SED, and WFSS measurements. We caution that individual values need not agree because the measurements differ in spectral resolution, wavelength coverage, extraction aperture, and sensitivity. WFSS H$\alpha$ is used only for validation and does not replace the PRISM or SED H$\alpha$ luminosity paired with Ly$\alpha$. Aperture and slit-loss uncertainties are discussed in Section~\ref{sec:data_sed}.

\section{Robustness of the Visibility--Environment Relation}
\label{app:visibility_robustness}

\begin{figure*}[!ht]
    \centering
    \includegraphics[width=\textwidth]{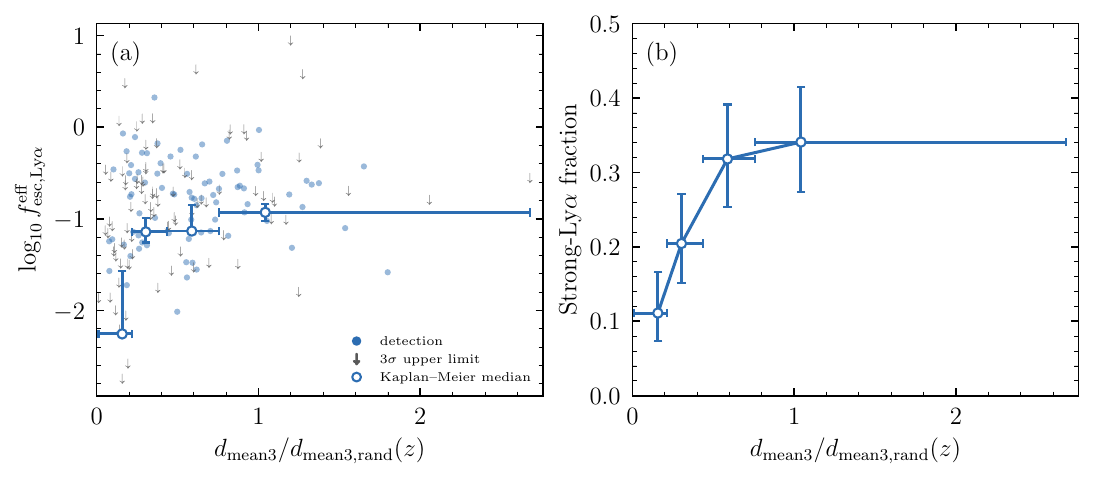}
    \caption{
    Ly$\alpha$ visibility as a function of normalized neighbor distance for the full environment sample. \textbf{(a)} Individual $f_{\rm esc,Ly\alpha}^{\rm eff}$ measurements and 3$\sigma$ upper limits, using one G140M or PRISM measurement per galaxy. Open circles show Kaplan--Meier medians in four equal-count environment intervals. Horizontal bars mark the interval boundaries, and vertical bars show the 16th--84th percentile ranges from resampling the galaxies. \textbf{(b)} Strong-Ly$\alpha$ fraction in the same intervals, with 68\% intervals. Smaller normalized distance corresponds to higher projected line-emitter density.
    }
    \label{fig:full_visibility_robustness}
\end{figure*}

\begin{figure*}[!ht]
    \centering
    \includegraphics[width=\textwidth]{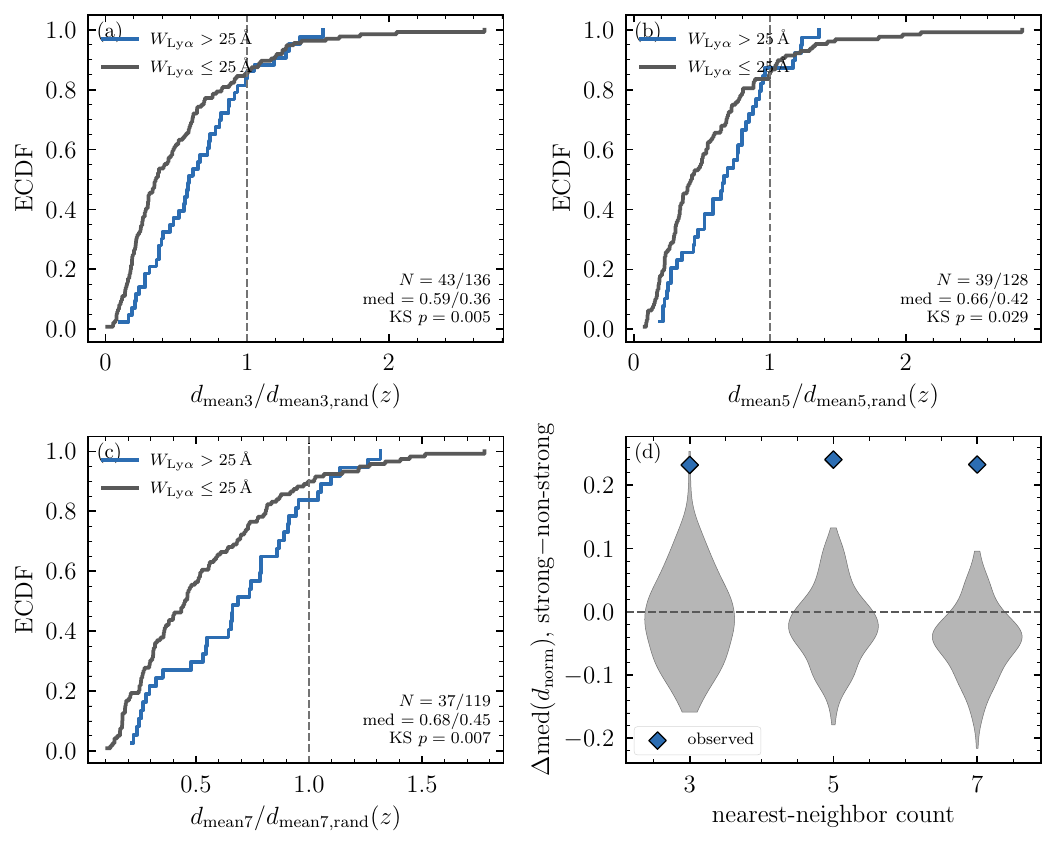}
    \caption{
    Dependence of the visibility--environment relation on the number of neighboring tracers. \textbf{(a)}--\textbf{(c)} Normalized-distance distributions for strong and non-strong Ly$\alpha$ emitters using the nearest 3, 5, and 7 WFSS tracers. The strong-minus-non-strong median separations are 0.232, 0.240, and 0.233, respectively. \textbf{(d)} Observed separations compared with those obtained after randomizing the tracer positions within the survey footprint. For the nearest-three statistic, one of 200 randomized realizations matches or exceeds the observed absolute separation.
    }
    \label{fig:neighbor_random_robustness}
\end{figure*}

\begin{figure*}[!ht]
    \centering
    \includegraphics[width=\textwidth]{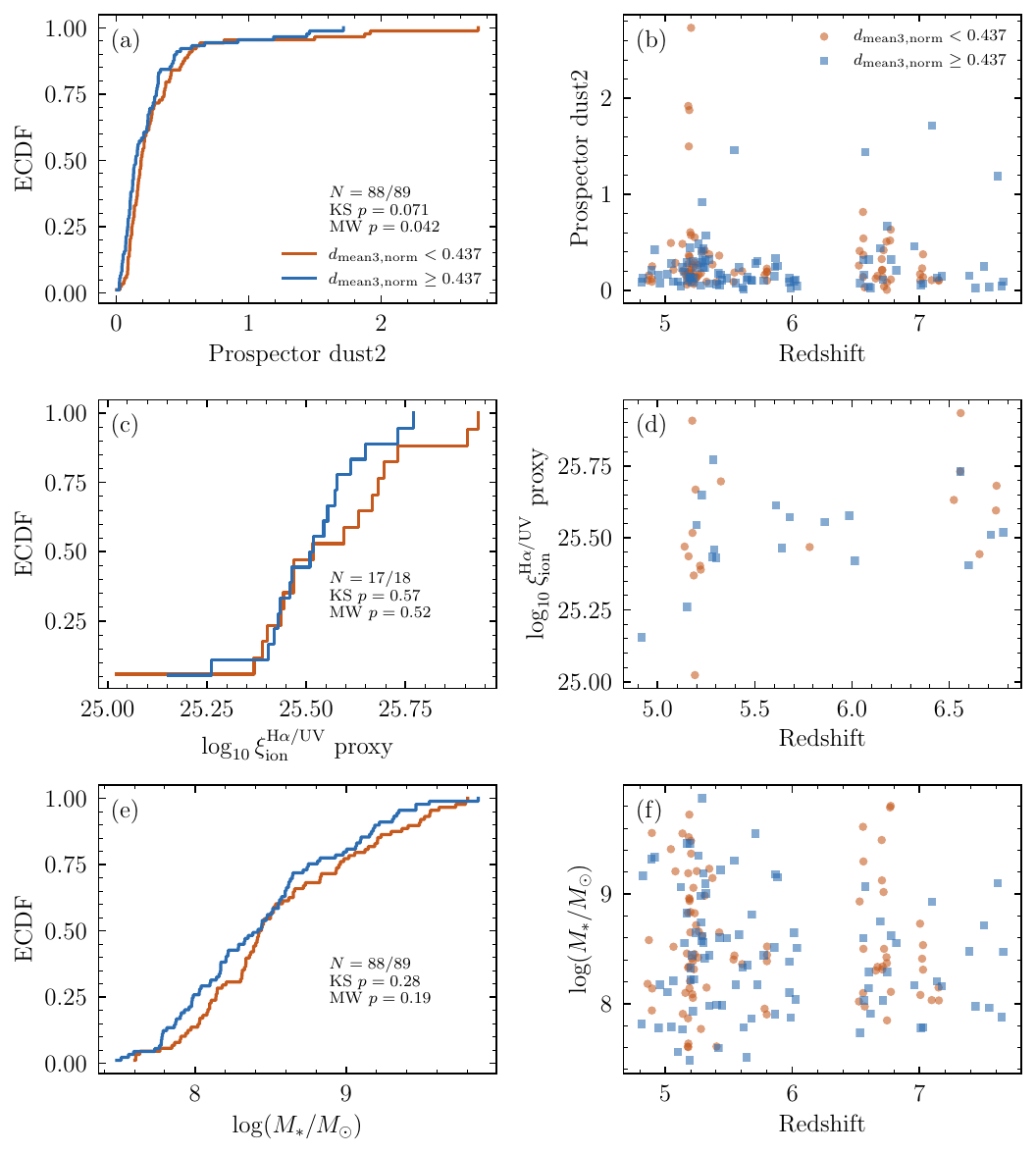}
    \caption{
    Galaxy properties as a function of normalized environment. The left column compares galaxies in denser and less dense line-emitter environments, while the right column shows the same measurements as a function of redshift. The rows show \textbf{(a,b)} Prospector dust attenuation, \textbf{(c,d)} the PRISM H$\alpha$/UV ionizing-production proxy, and \textbf{(e,f)} stellar mass.
    }
    \label{fig:galaxy_property_density_appendix}
\end{figure*}

Figure~\ref{fig:full_visibility_robustness} shows the individual $f_{\rm esc,Ly\alpha}^{\rm eff}$ measurements and upper limits as a function of normalized neighbor distance. The sample contains 82 detections and 95 upper limits. The Kaplan--Meier calculation includes both when estimating the visibility distribution. The median $\log_{10}f_{\rm esc,Ly\alpha}^{\rm eff}$ increases from $-2.26$ in the densest interval to $-0.93$ in the least dense interval. A log-rank comparison, which tests whether the visibility distributions differ while retaining the upper limits, gives $p=0.0044$. A second test for a steady change across the four environment intervals gives $p=2.5\times10^{-4}$. The strong-Ly$\alpha$ fraction independently increases from 0.11 to 0.34 over the same range.

We next test whether the result depends on the number of neighboring tracers used to define environment. Figure~\ref{fig:neighbor_random_robustness} repeats the strong/non-strong comparison using the mean projected distance to the nearest 3, 5, and 7 tracers. All three neighbor counts give nearly the same median separation and retain the result that strong Ly$\alpha$ emitters lie at larger normalized distances. We also randomize the tracer angular positions while preserving their redshifts. Only one of 200 randomized realizations produces a separation at least as large as the observed value, corresponding to $p=2/201=0.00995$. We retain the nearest-three statistic as the main environment measurement.

We caution that the normalization from random positions accounts for the survey boundary and the redshift-dependent number of tracers, but not for spatial variations in WFSS sensitivity or contamination. The catalog does not provide a continuous sensitivity map that could be used to correct these variations.

Because the non-detections are upper limits, we reverse the sign of $\log_{10}f_{\rm esc,Ly\alpha}^{\rm eff}$ before applying the standard Kaplan--Meier and log-rank calculations, which are formulated for limits in the opposite direction. The sign is changed back when reporting the results. When resampling the galaxies, overlapping G140M and PRISM measurements of the same source are kept together. For the randomization tests, the environment measurements are reassigned among galaxies while keeping the G140M and PRISM samples separate. Each galaxy contributes one measurement to the four intervals in Figure~\ref{fig:full_visibility_robustness}.

Finally, we test whether other measured galaxy properties show the same environmental separation as Ly$\alpha$. Figure~\ref{fig:galaxy_property_density_appendix} compares dust attenuation, stellar mass, and the H$\alpha$/UV ionizing-production proxy between galaxies above and below the median normalized neighbor distance. The dust distributions show a marginal difference, with KS $p=0.071$ and Mann--Whitney $p=0.042$, but the two tests do not give consistent evidence for a separation. The stellar-mass comparison gives KS $p=0.284$ and Mann--Whitney $p=0.187$, while the H$\alpha$/UV comparison gives KS $p=0.565$ and Mann--Whitney $p=0.520$. None of these quantities shows a separation as strong or as consistently supported as the Ly$\alpha$ result in Figure~\ref{fig:normalized_neighbor_escape}. We caution that these tests do not exclude contributions from star-formation rate, metallicity, dust geometry, star-formation history, the intrinsic Ly$\alpha$ profile, or other galaxy properties.
We also directly compare the Prospector \texttt{dust2} parameter for strong and non-strong Ly$\alpha$ emitters. The median values are 0.111 and 0.205, respectively, for 44 strong and 206 non-strong emitters.
The distributions differ with KS $p=0.0028$ and Mann--Whitney $p=0.00037$. Strong Ly$\alpha$ emitters therefore tend to have lower dust attenuation, consistent with dust contributing to Ly$\alpha$ visibility, although this does not show that dust drives the environmental trend.

\section{\texorpdfstring{Foreground Ly$\alpha$ Forest Transmission}{Foreground Lyalpha Forest Transmission}}
\label{app:tau_checks}

\begin{figure*}[!ht]
    \centering
    \includegraphics[width=0.92\textwidth]{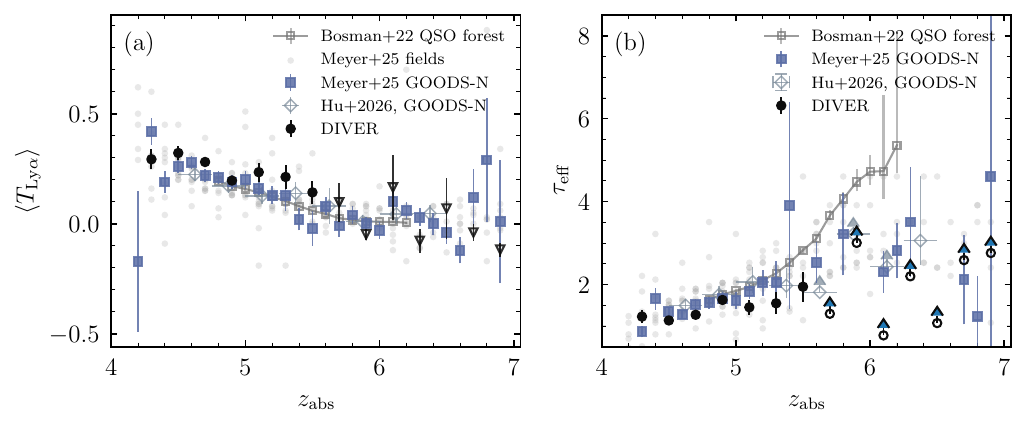}
    \caption{
    Foreground Ly$\alpha$ forest transmission measured from the DIVER PRISM spectra. \textbf{(a)} Mean normalized transmission as a function of absorber redshift. Gray symbols show the quasar-forest measurements from \citet{bosman_hydrogen_2022}. Light gray points show the JWST field measurements from \citet{meyer_probing_2025}, with their GOODS-N measurements highlighted in blue. Independent GOODS-N measurements from \citet{hu_using_2026} are also shown. \textbf{(b)} Corresponding effective optical depths. Positive transmission measurements are converted using $\tau_{\rm eff}=-\ln\langle T\rangle$; upward arrows mark lower limits where the transmission is consistent with zero.
    }
    \label{fig:tau_vs_z_appendix}
\end{figure*}

\begin{figure*}[!ht]
    \centering
    \includegraphics[width=\textwidth]{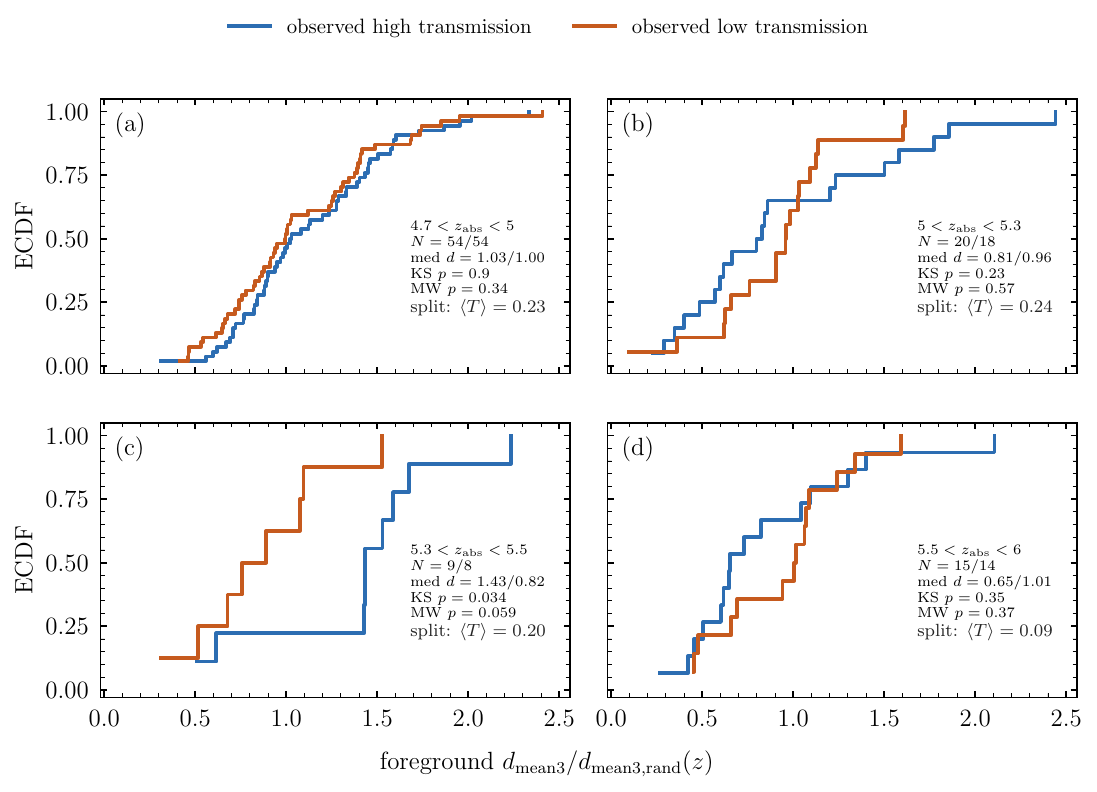}
    \caption{
    Foreground Ly$\alpha$ forest transmission as a function of foreground line-emitter environment. \textbf{(a)}--\textbf{(d)} Normalized-distance distributions for high- and low-transmission measurements in four absorber-redshift intervals. Forest pixels are grouped by galaxy, $\Delta z_{\rm abs}=0.1$ interval, and foreground environment before being averaged. High and low transmission are defined relative to the median transmission in each redshift interval. The clearest separation occurs at $5.3<z_{\rm abs}<5.5$.
    }
    \label{fig:topology_comparison}
\end{figure*}

As a supporting check, we measure foreground Ly$\alpha$ forest transmission in the DIVER PRISM spectra following \citet{meyer_probing_2025}. We fit BAGPIPES continua without IGM absorption to the red side of each spectrum and interpolate the models onto the observed wavelength grid. Forest pixels are selected over $1036<\lambda_{\rm rest}<1180$~\AA\ and assigned absorber redshifts using $z_{\rm abs}=\lambda_{\rm obs}/1215.67-1$. We then calculate the inverse-variance-weighted mean transmission in absorber-redshift intervals, with uncertainties estimated by resampling the spectra. Figure~\ref{fig:tau_vs_z_appendix} compares the DIVER measurements with previous galaxy- and quasar-forest measurements. For significant positive transmission, we calculate $\tau_{\rm eff}=-\ln\langle T\rangle$. When the transmission is consistent with zero, we report a lower limit on $\tau_{\rm eff}$. The DIVER measurements lie within the field-to-field scatter reported by \citet{meyer_probing_2025} and are broadly consistent with the independent GOODS-N measurements from \citet{hu_using_2026} over the overlapping redshift range.

Figure~\ref{fig:topology_comparison} compares the foreground transmission with the WFSS environment measurement. Forest pixels are grouped by galaxy, $\Delta z_{\rm abs}=0.1$ interval, and foreground environment before being averaged. Within each absorber-redshift interval, the measurements are divided at the median transmission. The high/low sample sizes are 54/54, 20/18, 9/8, and 15/14 from low to high redshift.
Only the $5.3<z_{\rm abs}<5.5$ interval shows evidence for a difference, with KS $p=0.0336$.
We caution that this is one of four redshift intervals. The PRISM resolution, continuum uncertainty, and small samples also limit the comparison. We therefore use the forest measurement as a supporting check on the galaxy Ly$\alpha$ visibility result.

\section{Additional Simulation Comparisons}
\label{app:simulation_comparisons}

\begin{figure*}[!ht]
    \centering
    \includegraphics[width=\textwidth]{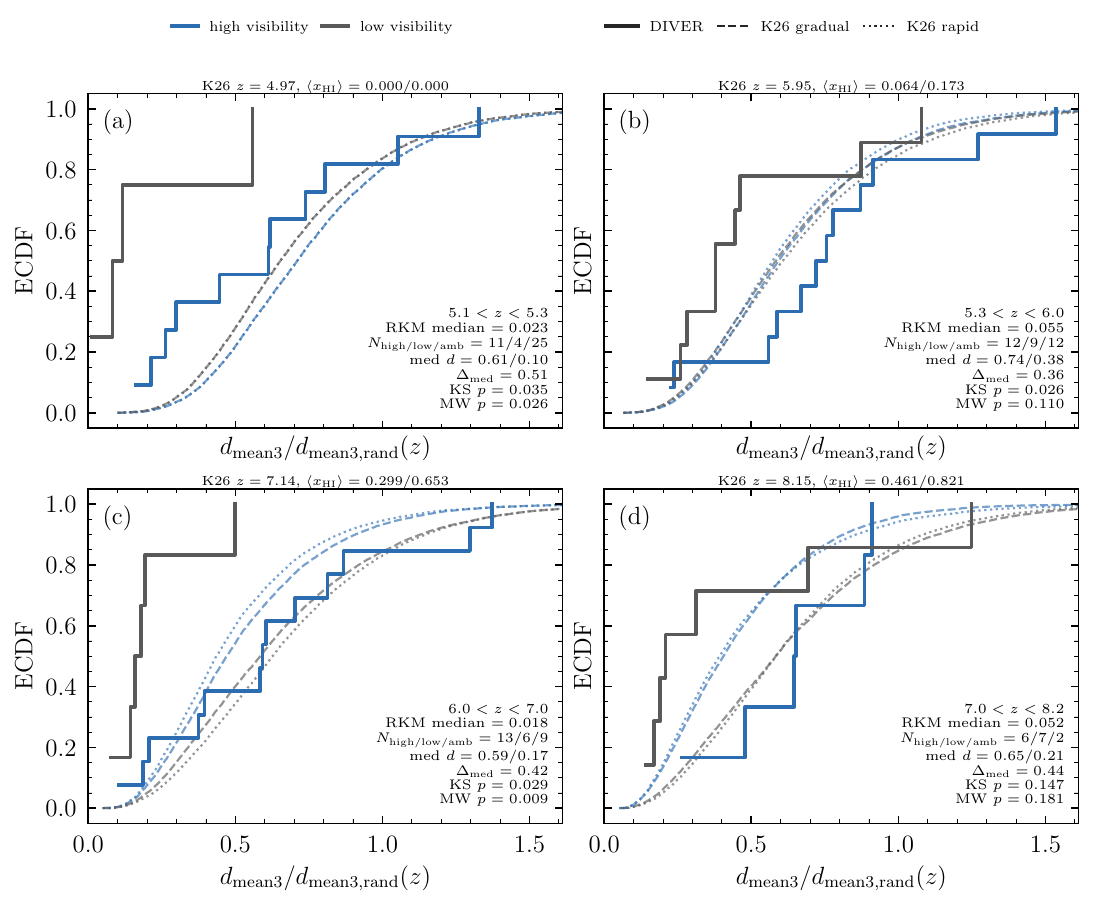}
    \caption{
    DIVER and K26 divided at the median Ly$\alpha$ visibility in each redshift interval. Solid curves show DIVER, while dashed and dotted curves show the K26 gradual and rapid histories. Blue and gray denote higher and lower visibility. The observational medians include both detections and upper limits. Upper limits that do not determine which side of the median a galaxy occupies are omitted from the two-sample comparison. Each K26 snapshot is divided at its median $f_{\rm esc,IGM}$. The comparison uses the redshift intervals from Figure~\ref{fig:normalized_neighbor_escape}.
    }
    \label{fig:grating_median_split_simulation}
\end{figure*}

\begin{figure*}[!ht]
    \centering
    \includegraphics[width=0.95\textwidth]{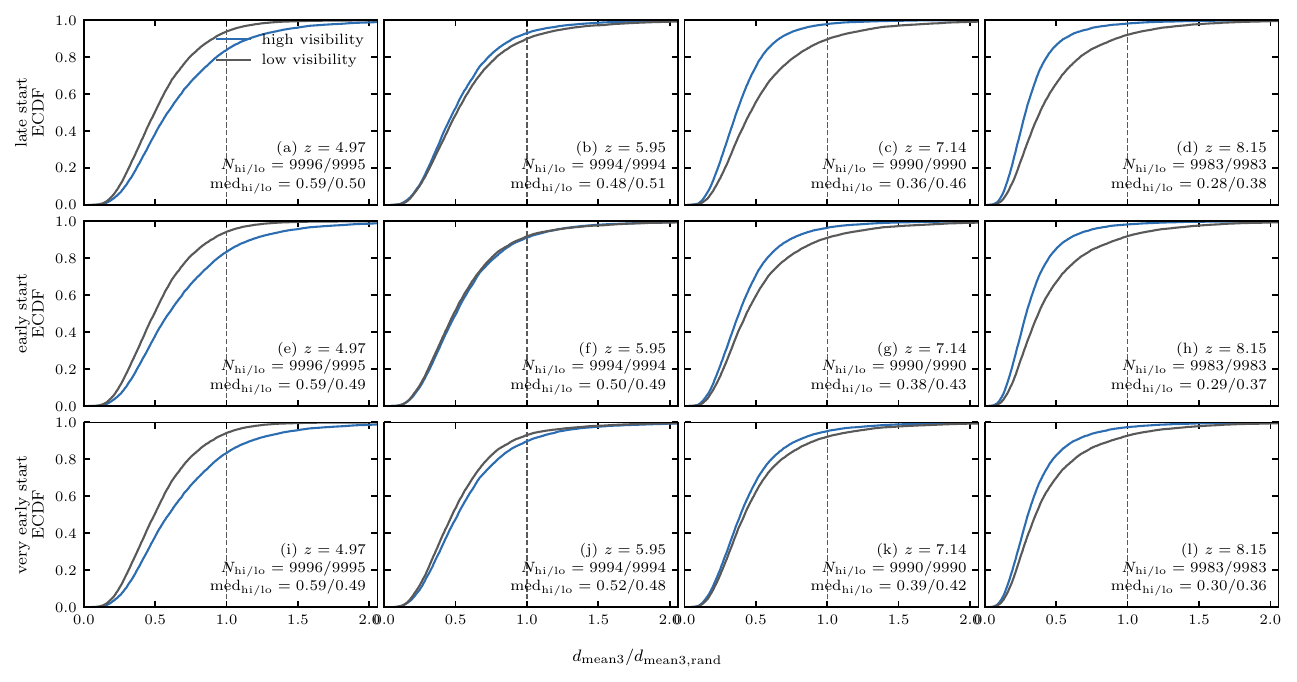}
    \caption{
    Ly$\alpha$ visibility and environment in the FlexRT reionization histories. Sources are ranked using the damping-wing opacity measured from the modeled spectra. Blue and gray curves show the higher- and lower-visibility subsets, and the vertical dashed line marks $d_{\rm mean3}/d_{\rm mean3,rand}=1$. Each panel lists the redshift, sample sizes, and median normalized neighbor distances.
    }
    \label{fig:flexrt_simulation_context}
\end{figure*}

Figure~\ref{fig:grating_median_split_simulation} compares DIVER and K26 after dividing each redshift interval into galaxies with higher and lower Ly$\alpha$ visibility. For DIVER, the Kaplan--Meier medians are 0.0229, 0.0552, 0.0176, and 0.0524 from the lowest to the highest redshift interval. The numbers above, below, and unconstrained relative to these medians are 11/4/25, 12/9/12, 13/6/9, and 6/7/2. The first interval has only four galaxies in the low-visibility sample, so its comparison is less certain. The high-minus-low median normalized-distance separations are 0.51, 0.36, 0.42, and 0.44. The corresponding K26 separations are smaller and become negative at higher redshift, as discussed in Section~\ref{sec:discussion_sims}. Dividing each sample at its own median allows us to compare the environmental ordering without treating $f_{\rm esc,Ly\alpha}^{\rm eff}$ and $f_{\rm esc,IGM}$ as the same quantity.

We also compare the DIVER result with the FlexRT late-start, early-start, and very-early-start reionization histories \citep{cain_flexrt_2024,cain_chasing_2025}. All three histories end reionization near $z\sim5$. The available outputs do not provide $f_{\rm esc,IGM}$, so we use the damping-wing opacity measured from the modeled spectra to rank sources by relative Ly$\alpha$ visibility. Figure~\ref{fig:flexrt_simulation_context} shows the same broad evolution as K26. At higher neutral fractions, sources with higher Ly$\alpha$ visibility tend to have smaller normalized neighbor distances, consistent with ionized regions growing around clustered sources. At lower neutral fractions, the separation weakens or reverses as density-dependent opacity becomes more important.

The FlexRT and K26 comparisons both show that the global neutral fraction alone does not determine the visibility--environment relation. CGM and local-IGM opacity, high-column-density absorbers, source selection, and galaxy properties may also contribute to the DIVER trend. We caution that the simulated sources and tracers are selected from halo catalogs and do not reproduce the DIVER, FRESCO, or CONGRESS line-flux selections. Differences in halo mass, duty cycle, line luminosity, and detectability can change the sampled environments. We therefore compare the sign and redshift evolution of the relation, not its absolute amplitude.

\clearpage

\input{IGM8018.bbl}
\end{document}

%% file: source_measurement_summary.tex
\begin{deluxetable*}{rrrrrrrrrrrrrrrrrr}[!ht]
\tablecaption{Source Measurement Catalog \label{tab:source_measurement_summary}}
\setlength{\tabcolsep}{1pt}
\tabletypesize{\tiny}
\tablewidth{0pt}
\tablehead{
\colhead{\#} &
\colhead{ID} &
\colhead{RA} &
\colhead{Dec} &
\colhead{$z$} &
\colhead{EW25} &
\colhead{$\log L_{\rm Ly\alpha}^{\rm G140M}$} &
\colhead{$W_{\rm Ly\alpha}^{\rm G140M}$} &
\colhead{$\log L_{\rm Ly\alpha}^{\rm PRISM}$} &
\colhead{$W_{\rm Ly\alpha}^{\rm PRISM}$} &
\colhead{$\log L_{\rm H\alpha}^{\rm PRISM}$} &
\colhead{$\log L_{\rm H\alpha}^{\rm SED}$} &
\colhead{$\log L_{\rm H\alpha}^{\rm WFSS}$} &
\colhead{$\log L_{\rm H\alpha}^{\rm adopt}$} &
\colhead{$f_{\rm esc,Ly\alpha}^{\rm eff}$} &
\colhead{$d_{\rm mean,3}$} &
\colhead{$d_{\rm mean,3}^{\rm rand}$} &
\colhead{$d_{\rm mean,3}/d_{\rm mean,3}^{\rm rand}$}
}
\startdata
1 & 1066572 & 189.242050 & 62.237363 & 4.8153 & Y &  &  & $42.46\pm0.10$ & $130.4\pm33.2$ & $41.93\pm0.03$ & $41.86\pm0.40$ & $41.80\pm0.06$ & $41.93\pm0.03$ & $0.389\pm0.097$ & 4.434 & 4.460 & 0.994 \\
2 & 1034159 & 189.114290 & 62.293003 & 4.8256 &  & $<42.51$ & $<120.7$ &  &  &  & $40.99\pm0.05$ & $41.64\pm0.07$ & $40.99\pm0.05$ & $<3.787$ & 5.678 & 4.460 & 1.273 \\
3 & 1115128 & 189.139639 & 62.241427 & 4.8365 &  &  &  & $<42.18$ & $<103.1$ & $<41.12$ & $42.34\pm0.28$ &  & $42.34\pm0.28$ & $<0.080$ &  &  &  \\
4 & 1018491 & 189.188368 & 62.286328 & 4.8374 &  & $<42.16$ & $<166.8$ & $<42.04$ & $<133.7$ &  & $41.65\pm0.38$ & $41.61\pm0.04$ & $41.65\pm0.38$ & $<0.286$ &  &  &  \\
5 & 1022983 & 189.193393 & 62.303268 & 4.8619 & Y & $42.23\pm0.10$ & $131.6\pm33.5$ &  &  &  & $41.95\pm0.39$ & $41.89\pm0.05$ & $41.95\pm0.39$ & $0.218\pm0.203$ & 1.800 & 4.460 & 0.404 \\
6 & 1010521 & 189.150570 & 62.258150 & 4.8722 &  & $<42.68$ & $<290.3$ &  &  &  & $41.65\pm0.33$ &  & $41.65\pm0.33$ & $<1.250$ & 1.542 & 4.460 & 0.346 \\
7 & 1012721 & 189.114403 & 62.267043 & 4.8850 &  &  &  &  &  & $41.94\pm0.02$ & $41.69\pm0.39$ & $41.72\pm0.04$ & $41.94\pm0.02$ &  &  &  &  \\
8 & 1020713 & 189.108300 & 62.293200 & 4.8923 &  & $<41.93$ & $<16.9$ &  &  &  & $42.52\pm0.33$ & $42.29\pm0.03$ & $42.52\pm0.33$ & $<0.029$ & 2.671 & 4.460 & 0.599 \\
9 & 1009935 & 189.098162 & 62.255551 & 4.8955 &  & $<42.26$ & $<21.4$ &  &  &  & $41.22\pm0.03$ & $42.28\pm0.02$ & $41.22\pm0.03$ & $<1.237$ & 1.256 & 4.460 & 0.282 \\
10 & 1003589 & 189.157193 & 62.235467 & 4.8956 & Y &  &  & $42.59\pm0.11$ & $268.2\pm76.0$ & $41.92\pm0.05$ & $41.91\pm0.39$ & $41.80\pm0.04$ & $41.92\pm0.05$ & $0.527\pm0.143$ & 1.251 & 4.460 & 0.280 \\
\enddata
\tablecomments{RA and Dec are in degrees, Ly$\alpha$ equivalent widths are rest-frame values in \AA, and luminosities are reported as $\log_{10}(L/{\rm erg\,s^{-1}})$. Quoted uncertainties are 1$\sigma$. The H$\alpha$ luminosities are observed values and are not corrected for dust attenuation. The column $L_{\rm H\alpha}^{\rm adopt}$ gives the H$\alpha$ luminosity used to calculate $f_{\rm esc,Ly\alpha}^{\rm eff}$: the PRISM measurement for a PRISM Ly$\alpha$ measurement and the SED estimate for a G140M Ly$\alpha$ measurement. The WFSS H$\alpha$ luminosity is used only for the consistency check in Appendix~\ref{app:data_validation}. EW25 marks galaxies with $W_{\rm Ly\alpha}>25$~\AA. For non-detections, the listed Ly$\alpha$ quantities and $f_{\rm esc,Ly\alpha}^{\rm eff}$ are 3$\sigma$ upper limits where available. The distances $d_{\rm mean,3}$ and $d_{\rm mean,3}^{\rm rand}$ are in cMpc, and their ratio is dimensionless. Blank entries indicate unavailable measurements. \\The complete 261-row table is available in machine-readable form.}
\end{deluxetable*}